\documentclass[reprint,amsmath,amssymb,nofootinbib]{revtex4-2}
\usepackage{graphicx}
\usepackage{xcolor}
\usepackage{amsthm}
\usepackage{amssymb}
\usepackage{stmaryrd}
\usepackage[utf8]{inputenc}

\usepackage{chemfig}
\usepackage{comment}
\usepackage{hyperref}
\usepackage{float}
\usepackage[normalem]{ulem}

\hypersetup{colorlinks = true, linkcolor=blue, citecolor=orange, urlcolor=blue, bookmarksnumbered = true}

\begin{document}

\title{Advanced approach of superconducting gap function extraction from tunneling experiments}
\author{František Herman}
\email{herman2@uniba.sk}
\affiliation{Department of Experimental Physics, Comenius University, Mlynská Dolina F2, 842 48 Bratislava, Slovakia}

\begin{abstract}
    An advanced theoretical framework is introduced and examined. Its main idea is to extract properties of the superconducting pairing gap function $\Delta(\omega)$ in the conventional, nearly localized superconductors. To test the approach, we present an experimentally relevant benchmark model with defined normal and superconducting sectors. The developed reverse engineering framework consists of two logic steps. First, dismantle the superconducting density of states into the effects coming from the superconducting pairing and effects inherited from the normal state. Second, extract and reconstruct properties of $\Delta(\omega)$ and compare it to the superconducting sector of the defined benchmark model. Applying this approach, we can: (i) simulate extraction from the actual experimental low-temperature tunneling data and comment on their required properties, and (ii) maintain absolute control above the reconstructed Cooper-pair-influencing properties during ameliorating the individual steps of the method.
\end{abstract}

\maketitle

{\it Introduction.} It has long been known that the density of states and its resulting low-temperature tunneling conductance of the conventional $s$-wave superconductors are dramatically different inside and outside of the region of the spectroscopic gap \cite{Tinkham}. The boundary region itself can usually be recognized by the (easily distinguishable) end of the gap and the raising of the coherence peak. Described properties are also very resilient against the disorder caused by the, often present, Cooper-pair-conserving electron scattering \cite{Anderson59, Herman16}. Naturally, scanning tunneling spectroscopy (STS) measurements are therefore very often used as standard experimental probes of the superconducting state \cite{Sacepe11, Szabo16, Groll18, Zemlicka20, Postolova20}.

Our main aim will therefore be the reconstruction of the gap function $\Delta(\omega)$ out of data for the density of states (dos\footnote{This function corresponds to the normalized superconducting density of states of the material with the same $\Delta(\omega)$, but with ideal, the constant density of state in the normal state.}) function $n(\omega)$ considering conventional low-$T_c$ superconductors, where \cite{Herman16}:
\begin{equation}
n(\omega) = {\rm Re}\left[\frac{\omega}{\sqrt{\omega^2-\Delta(\omega)^2}}\right].
\end{equation}
$\Delta(\omega)$ representing the gap function contains the information about the electron pairing interactions.

Routinely, the standard procedure to interpret STS data of the disordered superconductors commences with introducing the model of the underlying density of states of the studied sample material. Next, the assumed model is used to fit the actual STS data. If we focus on the disordered superconductors using the Dynes formula as the model for the superconducting density of states \cite{Szabo16, Dynes78, Herman16, Groll18}, the resulting spectroscopic information describing the superconducting state usually consists of three characteristic energy scales. The first corresponds to the considered temperature. The other two specify the value of the pair-creating superconducting gap $\Delta$ and the pair-breaking scattering rate $\Gamma$. Actual values of $\Delta$ and $\Gamma$ result from the fitting procedure using the Dynes formula for the tunneling density of states in the superconducting state. Eventually, encountering the non-constant density of states in the normal state requires enriching the assumed parameters describing the normal state, as in Ref.~\cite{Dynes86, Zemlicka20, Postolova20, Carbillet20} and references therein.

Despite the described recipe being very successful in interpreting the main characteristics and the overall trend of the corresponding STS data, it might overlook essential details hidden in their minutiae. After all, focusing on these more detailed imprints \cite{Dynes86, Zemlicka20, Postolova20, Carbillet20, Roy19, Meservey70, Noat13, Wu96, Butko99, Jujo19}, can allow us to understand and describe normal and superconducting states in more detail than before.

One could therefore ask a very natural question: "Could we process the tunneling data in such a way, that distinguishes and describes the effects of superconducting and normal state without prior model assumption?" Or, since we focus primarily on the superconducting state: "Could we reveal the full information about the pair-creating \cite{McMillan65} and the pair-breaking \cite{Bzdusek15, Herman16, Kavicky20} properties of the superconducting state with no, or at least absolutely minimal, prior model assumption about the underlying density of states?" If we would answer yes to these or similar questions, we could go beyond the current status of knowledge. Elaborating and answering these and similar questions represents the very core idea of our manuscript.

Pioneering work in this direction was introduced recently in the Ref.~\cite{Kavicky22}. This work represents a natural continuation of our previous effort to extract the gap function of nearly localized conventional superconductors directly from the experimental data. In this manuscript, an alternative approach is presented. 

Compared with the Ref.~\cite{Kavicky22}, we derive the fundamental starting point equations in a more general way. Relaxing one of the assumptions allows us to obtain more complete equations relating the odd and even parts of superconducting and normal densities of states. In this manuscript, the nonzero odd parts of densities of states in the normal state are also assumed and exploited. Of course, this discussion in principle holds true also to clean superconductors and the introduced method can be applied also to them.

However, as an example, where the reconstruction of the gap function can be helpful, let us mention the increasing impact of the STS measurements on the research area of the quantum breakdown of superconductivity (QBS) \cite{Sacepe08, Sacepe10, Mondal11, Chand12, Sherman12, Kamlapure13, Ganguly17, Zhao19, Lemarie13, Carbillet16, Carbillet20}, revealing emergent granularity on the local scale \cite{Kamlapure13, Ganguly17}. In principle, the gap function reconstruction can help to distinguish the roles of separate scenarios of the QBS \cite{Goldman98, Sacepe20}, induced by the increasing disorder \cite{Sacepe20, Anderson83, Finkelstein94, Doniach81, Fisher90, Ma85, Feigelman10}. Knowledge of $\Delta(\omega)$ helps us to identify the role of the so-called fermionic (amplitude of the order parameter driven) scenario \cite{Anderson83, Finkelstein94}, within the conundrum of the QBS transition. In this way, we would be able to recognize the fundamental nature of the charge carriers on both sides of the considered quantum phase transition. In the pure fermionic scenario, the normal state consists of electrons, meanwhile, in the pure bosonic (phase of the order parameter driven) scenario we have non-communicating Cooper pairs.

Since the method which we are going to introduce requires detailed explanation and testing, we structure this manuscript in the following way: First, we describe the theory and the necessary assumptions in Sec.~\ref{Sec:Theory}. Next, the detailed motivation and formulation of the problem are clarified in Sec.~\ref{Sec:Motiv}. Properties of the normal and the superconducting sectors of the benchmark model (BM) are defined and explained in Sec.~\ref{Sec:BM}. Individual steps of the numerical procedure itself demonstrated on the solution of the introduced problem are discussed in Sec.~\ref{Sec:Solution}. 

For further clarification, within Sec.~\ref{Sec:Solution} we provide a test of the developed numerical procedure on the BM example, comparing it to the structure known from its original microscopic construction. In this way, we will be able to quantify and show how accurately we can reconstruct the gap function $\Delta(\omega)$. Technical details, limiting cases, and more complicated benchmark model (MCBM) example can be found in the Appendixes:~\ref{Appendix:RatFun}, \ref{Appendix:MandTech}, \ref{Appendix:Gap_in_infty},  \ref{Appendix:MCBM}.

The experimental data and their processing most often bring an extra degree of challenges, that require a non-trivial explanation, before we start the extraction of the $\Delta(\omega)$ by itself. E.g.: the effect of the finite temperature smearing, averaging the measured current data to get smooth tunneling conductance data, normalization of the STS spectra etc.. Therefore, to progress thoroughly step-by-step, we decided to dedicate to such a work an extra attention in the future project. In this way, we can focus better on the various aspects of the discussed theory and techniques in this manuscript. We also avoid the scenario in which we overwhelm and confuse the reader with quanta of information focused on the processing of the experimental data. In the concluding part of this manuscript, we specify feasible requirements, which should allow the application of the introduced procedure of the gap function extraction on the actual experimental data.

\section{Theory}\label{Sec:Theory}

\subsection*{Spectral functions in the normal and superconducting state}
At first, we elaborate on the advantageous form of the general structure of the spectral functions, describing conventional, low-temperature superconductors \cite{Parks69a, Herman17a}. In order to proceed, it shows to be efficient to translate the information from two complex Eliashberg functions, describing wavefunction renormalization $Z(\omega)$ and the gap function $\Delta(\omega)$, to four real, auxiliary functions $\widetilde{\omega}(\omega),\, \widetilde{\gamma}(\omega),\, \widetilde{\Omega}(\omega),\, \widetilde{\Gamma}(\omega)$ by means of\footnote{We choose the branch of the square root in a way that the signs of $\widetilde{\Omega}$ and $\omega$ are the same. In this way, we can assume $\widetilde{\Gamma}>0$.}:
\begin{eqnarray}\label{eq:AuxialiaryFun}
Z(\omega) \omega&=&\widetilde{\omega}(\omega)+i\widetilde{\gamma}(\omega),
\nonumber
\\
Z(\omega)\sqrt{\omega^2-\Delta^2(\omega)}&=&\widetilde{\Omega}(\omega)+i\widetilde{\Gamma}(\omega).
\end{eqnarray}
Using these auxiliary functions, the spectral functions in the superconducting state can be elegantly written as \cite{Herman17a}:
\begin{multline}\label{eq:As}
A_s(\varepsilon,\omega) =
\frac{1}{2}\left(\frac{\widetilde{\gamma}}{\widetilde{\Gamma}}+1\right)\delta_{\widetilde{\Gamma}}(\varepsilon-\widetilde{\Omega})
+\frac{1}{2}\left(\frac{\widetilde{\gamma}}{\widetilde{\Gamma}}-1\right)\delta_{\widetilde{\Gamma}}(\varepsilon+\widetilde{\Omega})\\
+ \frac{1}{2}
\left(\frac{\widetilde{\omega}}{\widetilde{\Omega}}-\frac{\widetilde{\gamma}}{\widetilde{\Gamma}}\right)\frac{4\pi\widetilde{\Omega}^2}{\widetilde{\Gamma}}
\delta_{\widetilde{\Gamma}}(\varepsilon-\widetilde{\Omega})
\delta_{\widetilde{\Gamma}}(\varepsilon+\widetilde{\Omega}),
\end{multline}
where:
$$
    \delta_{\Gamma}(\Omega)=\frac{1}{\pi} \frac{\Gamma}{\Omega^2 + \Gamma^2}.
$$
Notice, that $\widetilde{\Omega}$ and $\widetilde{\Gamma}$ have natural interpretation of the quasiparticle energy and quasiparticle inverse lifetime respectively.

The dos function $n(\omega)$ and its Kramers-Kronig partner $\kappa(\omega)$ can be expressed as:
\begin{equation}\label{eq:comDOS}
\widetilde{n}(\omega) = n(\omega) + i\kappa(\omega) = \frac{\omega}{\sqrt{\omega^2-\Delta^2(\omega)}} = 
\frac{\widetilde{\omega}(\omega)+i\widetilde{\gamma}(\omega)}
{\widetilde{\Omega}(\omega)+i\widetilde{\Gamma}(\omega)}.
\end{equation}

One can also easily check the normal state limit of the spectral functions. We will assume that in such a case we obtain the solution of the Eq.~\eqref{eq:AuxialiaryFun} in the form:
$\Delta(\omega) = 0$, $\widetilde{\omega}(\omega)=\widetilde{\Omega}(\omega)=\omega$, $\widetilde{\gamma}(\omega)=\widetilde{\Gamma}(\omega)=\Gamma_n$, where $\Gamma_n$ is the normal state scattering constant, and so $Z(\omega) = 1+i\Gamma_n/\omega$. Combining this solution together with the Eq.~\eqref{eq:As},
leads to the well known Lorentzian form $A_{n}(\varepsilon, \omega) = \delta_{\Gamma_n}\left(\varepsilon - \omega\right)$.

\subsection*{Densities of states in the normal and superconducting state}

In what follows, we use the language of the superconducting (normal) spectral functions $A_{s,n}(\varepsilon,\omega)$, carrying the information about scattering and effects of the electron-phonon interaction. If we combine it with the density of states of the clean material in the normal state ${\cal N}_0(\varepsilon)$, we can express the superconducting (normal) state tunneling density of states $N_{s,n}(\omega)$ in the form \cite{Parks69b}:
\begin{eqnarray}\label{eq:Nsn}
N_{s,n}(\omega) = \int_{-\infty}^{\infty} d\varepsilon {\cal N}_0(\varepsilon) A_{s,n}(\varepsilon,\omega).
\end{eqnarray}

In order to take the advantage of the knowledge of the spectral functions, let us formally rewrite the density of tunneling states as a real part of the following contour integral on the closed contour $C$ in the upper half-plane of the complex space: 
\begin{eqnarray}\label{eq:Nsn}
N_{s,n}(\omega)={\rm Re}\oint_C dz \widetilde{{\cal N}_0}(z) A_{s,n}(z,\omega),
\end{eqnarray}
where we assume that the contour $C$ consists of a straight line, going from $-\infty$ to $\infty$, infinitesimally close above the real axis and semicircle, which closes the contour, with the radius growing to infinity. Since the $A_s(z,\omega)=A_n(z,\omega)\propto z^{-2}$, assuming $z\rightarrow\infty$ and we will construct $\widetilde{N_0}(z)$ to be analytic in the upper half-plane of the complex space, the semicircle contribution of the Eq.~\eqref{eq:Nsn} is zero. 

Since we want to use the knowledge of the position of poles of the available spectral functions $A_{s,n}(z,\omega)$, we have to consider the function $\widetilde{N_0}(z)$ to be analytic in the upper half-plane of the complex space. Therefore, we have to consider the density of states $N_0(z)$ together with its Kramers-Kronig partner $K_0(z)$, in the form:
\begin{align*}
\widetilde{{\cal N}_0}(z) &= N_0(z) + i K_0(z),\\
                &= N_{0,e}(z) + N_{0,o}(z) + i \big(K_{0,o}(z) + K_{0,e}(z)\big),\\
    K_0(y) &= \frac{1}{\pi}{\rm P} \int_{-\infty}^{\infty} d x \frac{N_0(x)}{y-x},
\end{align*}
where the even $N_{0,e}(z)$, $\left(K_{0,e}(z)\right)$ and odd $N_{0,o}(z)$, $\left(K_{0,o}(z)\right)$ parts of $N_0(\omega)$ and $K_0(\omega)$ can be related in the way:
\begin{align*}
    K_{0,o}(y) &= \frac{2y}{\pi}{\rm P} \int_{0}^{\infty} dx \frac{ N_{0,e}(x)}{y^2-x^2},\\
    K_{0,e}(y) &= \frac{2}{\pi}{\rm P} \int_{0}^{\infty} dx \frac{x N_{0,o}(x)}{x^2-y^2}.
\end{align*}
Next, we use Eq.~\eqref{eq:As} within the Eq.~\eqref{eq:Nsn}. Using Cauchy's integral formula, we are, after a few steps of cumbersome algebra, left with:
\begin{multline}\label{eq:Ns}
N_s(\omega) = \frac{1 + \widetilde{n}(\omega)}{2}\widetilde{{\cal N}_0}\big(\widetilde{\Omega}(\omega) + i\widetilde{\Gamma}(\omega)\big)\\
 \qquad -\frac{1 - \widetilde{n}^*(\omega)}{2}\widetilde{{\cal N}_0}\big(-\widetilde{\Omega}(\omega) + i\widetilde{\Gamma}(\omega)\big).
\end{multline}
For completeness, the situation in the normal state results in $N_n(\omega) = N_0(\omega + i\Gamma_n)={\rm Re}\,\widetilde{{\cal N}_0}(\omega + i\Gamma_n)$.

For further usage, we divide the Eq.~\eqref{eq:Ns} into the even and odd part:
\begin{align}\label{eq:Nseo}
N_{s,e}(\omega) &= n(\omega)N_{0,e}\big(\widetilde{\Omega}(\omega)+i\widetilde{\Gamma}(\omega)\big)\nonumber \\
&\qquad\qquad\qquad -\kappa(\omega)K_{0,o}\big(\widetilde{\Omega}(\omega)+i\widetilde{\Gamma}(\omega)\big),\nonumber\\
N_{s,o}(\omega) &= N_{0,o}\big(\widetilde{\Omega}(\omega)+i\widetilde{\Gamma}(\omega)\big).
\end{align}

\subsection*{Constant quasiparticle lifetime approximation}
In the next step, we focus on the situations in which we assume the so-called constant, $\widetilde{\Gamma}(\omega) = \Gamma_n$, approximation. In such a case, Eq.~\eqref{eq:Nseo} can be rewritten as:
\begin{align}
N_{s,e}(\omega) &\approx n(\omega)N_{0,e}\big(\widetilde{\Omega}(\omega)+i\Gamma_n\big)\nonumber\\
&\qquad\qquad\qquad -\kappa(\omega)K_{0,o}\big(\widetilde{\Omega}(\omega)+i\Gamma_n\big),\nonumber\\
&\approx n(\omega)N_{n,e}\big(\widetilde{\Omega}(\omega)\big)
-\kappa(\omega)K_{n,o}\big(\widetilde{\Omega}(\omega)\big),\label{eq:Nse_const_Gamma}\\
N_{s,o}(\omega) &\approx N_{0,o}\big(\widetilde{\Omega}(\omega)+i\Gamma_n\big),\nonumber\\
&\approx N_{n,o}\big(\widetilde{\Omega}(\omega)\big),
\label{eq:Nso_const_Gamma}
\end{align}
where we recognized even (odd) tunneling densities of states in the normal state $N_{n,e}$ ($N_{n,o}$), smeared by the normal state scattering rate. $K_{n,o}$ represents the Kramers-Kronig partner of $N_{n,e}$.

Even without the proper solution of the problem defined by Eqs.~\eqref{eq:Nse_const_Gamma}, and \eqref{eq:Nso_const_Gamma}, we can already notice its qualitatively interesting features. $N_{s,e}$ in the Eq.~\eqref{eq:Nse_const_Gamma} requires besides $\widetilde{\Omega}$ also knowledge of $\kappa$ and $K_{n,o}$, which are the Kramers-Kronig partners of $n$ and $N_{n,e}$ respectively. Typical qualitative behavior of the $\kappa K_{n,o}(\widetilde{\Omega})$ contribution shows, that this term will be localized in the region of coherence peaks \cite{Vega21, Vega22, Crespo06, Guillamon08}. Let us remind you that this region shows to be very important for discussing properties of the superconducting behavior \cite{Sacepe11, Kavicky22, Carbillet20, Feigelman7, Feigelman10, Feigelman12, Ghosal01}. Also, once $N_{n,e}(\omega)\neq const.$ in Eq.~\eqref{eq:Nse_const_Gamma}, $N_{s,e}(\omega)$ start to be dependent also on the pair-conserving scattering processes present in the overall scattering rate $\Gamma_n$. In cases, when $\Gamma_n \gtrsim $ "Energy scale of the developed Altshuler-Aronov minimum \cite{Altshuler79} in the normal state", we expect features of $N_n(\omega)$ be smeared to nearly a constant. To complete qualitative properties, Eq.~\eqref{eq:Nso_const_Gamma} can be also viewed as the implicit equation for the function $\widetilde{\Omega}(\omega)$.

Notice, that if we would now assume that the second term in the Eq.~\eqref{eq:Nse_const_Gamma} is negligible in comparison with the first one, and $n(\omega)=d\widetilde{\Omega}/d\omega$ due to the conservation of the total number of states, we would end up at the beginning of the approach developed and discussed in the Ref.~\cite{Kavicky22}. Recognizing known results in the limit case moves us further towards a different, more general approach. In what follows, we will keep also the second $\kappa K_{n,o}(\widetilde{\Omega})$ term and the only thing we will assume, besides $\widetilde{\Gamma}(\omega) = \Gamma_n$, is that $n(\omega) + i\kappa(\omega)$ is an unknown analytic function of $\omega$ that embodies the information about the gap function $\Delta (\omega)$.

\section{Motivation and Introduction of the Problem}\label{Sec:Motiv}

After the introduction of the theory background, we proceed further towards the development and implementation of the solution of the resulting Eqs.~\eqref{eq:Nse_const_Gamma} and \eqref{eq:Nso_const_Gamma} in the model example. The idea of this section is following. First, introduce some microscopically physically relevant model for $\Delta(\omega)$ (and $Z(\omega)$ for completeness) and calculate its experimentally available footprint in the form of the tunneling densities of states $N_{n,s}(\omega)$ in the normal and superconducting state. 

Afterward, take the resulting $N_{n,s}(\omega)$ as a result of the independent experiment and by solving Eq.~(\ref{eq:Nse_const_Gamma}, \ref{eq:Nso_const_Gamma}) reconstruct the behavior of the $\Delta(\omega)$. We will explain the details of this reverse engineering approach as we proceed. In this way, we gain the advantage of having under the control both, the microscopic input $\Delta(\omega)$ (resp. measurable $N_{n,s}(\omega)$), and also the output $N_{n,s}(\omega)$ (resp. $\Delta(\omega)$). We will also identify the required properties of the data for the future application of the method on the experimental data samples.

Let us remind you that we focus on the interpretation of the spectroscopic properties imprinted in the tunneling density of states. Therefore, as one can understand from the Eq.~\eqref{eq:comDOS}, we focus on the reconstruction of the gap function $\Delta(\omega)$, completely omitting the reconstruction of the $Z(\omega)$ for now.

\section{Definition of the theoretical Benchmark model}\label{Sec:BM}

\begin{figure}[htpb]
\includegraphics[width = 7.5 cm]{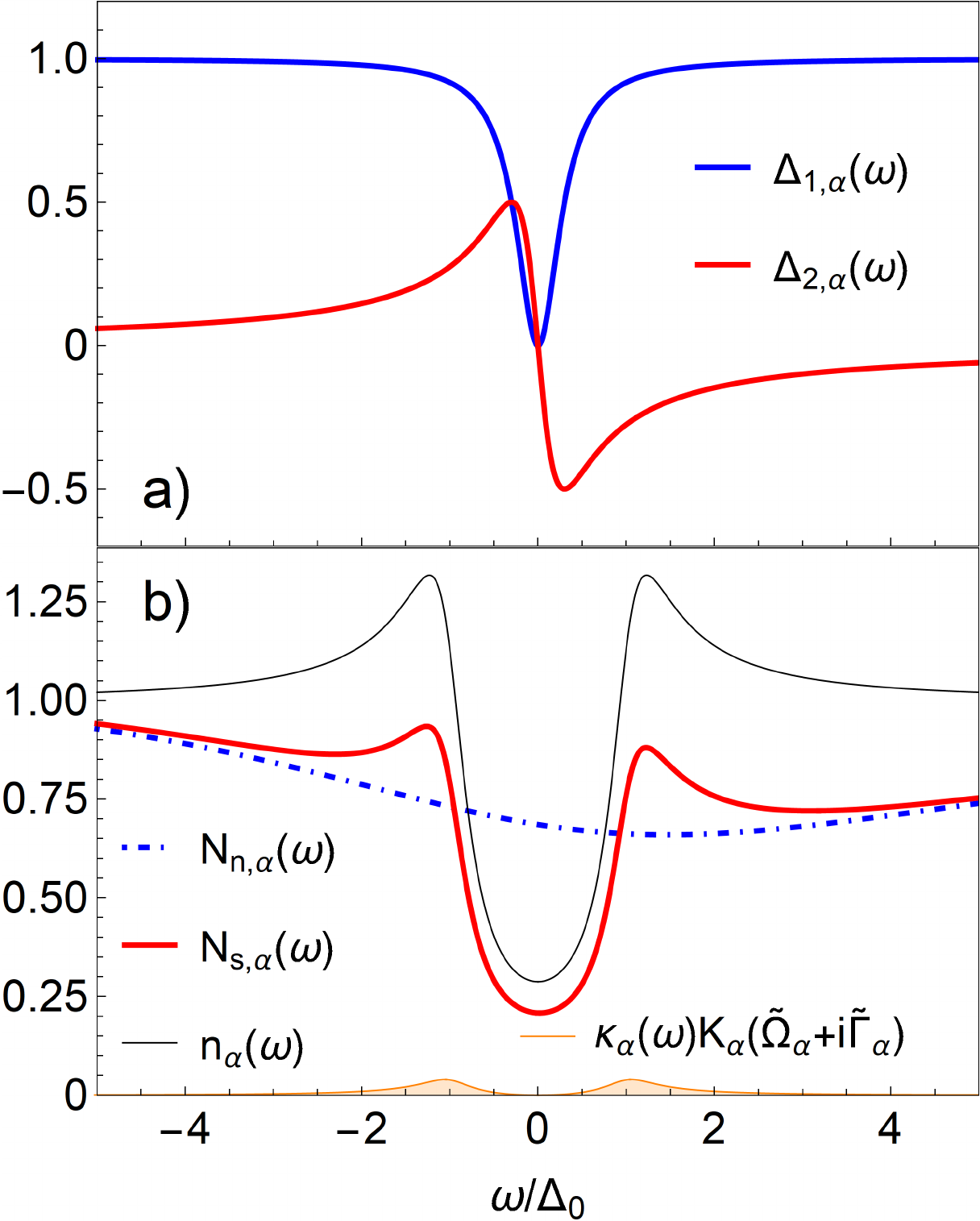}
\caption{Evaluation of the BM, considering values $\Gamma/\Delta_0 = 0.3$, and $\Gamma_s/\Delta_0 = 0$, $\alpha_e = -1/3$, $\Gamma_e/\Delta_0 = 5$, $\alpha_o = 0.2$ and $\Gamma_o/\Delta_0 = 5$.}
\label{Fig:BM}
\end{figure}

\subsection{Superconducting state}
As starting line, we model the properties of the superconducting state. The superconducting part of our benchmark model (BM) corresponds to the theory of the Dynes superconductor (DS) \cite{Herman16, Herman17a, Herman17b, Herman18, Herman21}. Our choice is based on the fact that the DS theory is a non-perturbative generalization of the Bardeen-Cooper-Schrieffer (BCS) theory. The superconducting pairing interaction of electrons into Cooper pairs is characterized by the value of the BCS gap $\Delta_0$. Within DS theory, we enrich superconducting pairing together with pair-breaking and pair-conserving scattering processes. These are represented by the presence of the scattering rates $\Gamma$ and $\Gamma_s$. The advantage of the DS theory is its simplicity (once formulated) and validity in different regimes comparing values of all involved energy scales characterized by $\Delta_0$, $\Gamma$, $\Gamma_s$, and $\omega$. Also, DS proves to be applicable to all sorts of sample materials, coming from various areas of research. For example, macroscopic $\chemfig{NbN}$ samples used in superconducting cavities applications: \cite{Herman21, Bafia21}, or single crystals of $\chemfig{La_2Ni_2In}$ \cite{Maiwald20}, which is interesting with regard to the formation of the quantum critical point (QCP), where, as already discussed earlier, the system is transitioning among different phases \cite{Sacepe20}.

If we want to consider the DS theory as the superconducting part of our BM we have to specify functions $\Delta(\omega)$ and $Z(\omega)$, introduced in the Eq.~\eqref{eq:AuxialiaryFun}, in the form presented in the Ref.~\cite{Herman16}:
\begin{eqnarray}\label{eq:BM_Delta_Z}
\Delta_{\alpha}(\omega) &=& \frac{\omega \Delta_0}{\omega + i\Gamma}, \nonumber\\
Z_{\alpha}(\omega) &=& \left(1 + \frac{i\Gamma}{\omega}\right)\left(1 + \frac{i\Gamma_s}{\sqrt{(\omega + i\Gamma)^2 - \Delta_0^2}}\right).
\end{eqnarray}
We plot the BM superconducting gap function $\Delta_{1,\alpha}(\omega)+i\Delta_{2,\alpha}(\omega)$ together with values of the considered parameters in the Fig.~\ref{Fig:BM}~a). Our main goal in the Sec.~\ref{Sec:Solution} will be the reconstruction of its behavior. Notice also, that the influence of nonzero $\Gamma_s$ would lead just towards a larger effect of smearing of the normal state density of states, since $\Gamma_n = \Gamma + \Gamma_s$, therefore, focusing on the most challenging case, we consider $\Gamma_s = 0$. 

\subsection{Normal state}
In order to reconstruct the superconducting properties of the underlying state and distinguish them from the normal state, we have to specify $N_{n,e}(\omega)$ (which also specifies its Kramers-Kronig partner $K_{n,o}(\omega)$) and $N_{n,o}(\omega)$ behavior in the Eq.~(\ref{eq:Nse_const_Gamma}, \ref{eq:Nso_const_Gamma}). In principle, we have only two requirements on the form of the density of state in the normal state. First is the local minimum of the tunneling density of states, caused by the plausible Altshuler-Aronov effect \cite{Altshuler79}. This effect could potentially disguise the properties of the superconducting gap and therefore, it has to be distinguished in the extraction process of the gap function. Our second requirement is $N_{n,o}(\omega)\neq 0$, since we also want to demonstrate the solution of the Eq.~\eqref{eq:Nso_const_Gamma} in order to reconstruct the function $\widetilde{\Omega}(\omega)$. Using the properties of the analytic rational complex functions $\rho_n(x)$, described in the Appendix~\ref{Appendix:RatFun}, we model the normal state properties in the form: 
\begin{eqnarray}
    \widetilde{N_0}_{,\alpha}(\omega) &=& 1 + \frac{\alpha_e}{1-i\omega/\Gamma_e} + \frac{i \alpha_o}{1-i\omega/\Gamma_o},\nonumber\\
    &=& 1 + \alpha_e \rho_0(\omega/\Gamma_e) + i \alpha_o \rho_0(\omega/\Gamma_o)\label{eq:BM_N_0}.
\end{eqnarray}
Nonzero parameters $\alpha_e$, $\alpha_o$, $\Gamma_e$, and $\Gamma_o$ specify the properties of the minimum and asymmetry around the Fermi energy in the normal state. We plot the BM normal $N_{n,\alpha}(\omega)$, BM superconducting $N_{s,\alpha}(\omega)$, and BM superconducting with constant normal $n_{\alpha}(\omega)$ densities of states using set of Eq.~(\ref{eq:comDOS}, \ref{eq:Nse_const_Gamma}, \ref{eq:Nso_const_Gamma}, \ref{eq:BM_Delta_Z}, \ref{eq:BM_N_0}) in the Fig.~\ref{Fig:BM}~b). We specify the values of the considered parameters in the caption of the same figure. We also highlight the amplitude of the original $\kappa_{\alpha} K_{\alpha,0,o}$ term from the Eq.~\eqref{eq:Nseo} by the orange color.

At the end of this section, let us remind all of the considered parameters in our particular model. Seven parameters of the BM are: $\Delta_0$, $\Gamma$, $\Gamma_s$, $\alpha_e$, $\Gamma_e$, $\alpha_o$, $\Gamma_o$. For simplicity, we use non-dimensional units $\Gamma_X/\Delta_0$, which means, that we have to specify $6$ numbers.

We also want to call attention to the Appendix~\ref{Appendix:MCBM}, where we introduce and subsequently reconstruct a more complicated model. However, for explanatory purposes of solving tasks ordered according to their difficulty level, we present in detail the BM model first.

\section{Solution}\label{Sec:Solution}

Throughout this section, we use the knowledge of the theory introduced in Sec.~\ref{Sec:Theory}. We provide step-by-step reverse engineering reconstruction of the superconducting gap function from the Eq.~\eqref{eq:BM_Delta_Z}. We use only data of the normal $N_{n,\alpha}(\omega)$ and superconducting densities of states $N_{s,\alpha}(\omega)$, generated in the previous section. 

\subsection{Solution for $\widetilde{\Omega}(\omega)$}\label{SubSec:Sol_Tilde_Omega}

\begin{figure}[h]
\includegraphics[width = 5.5 cm]{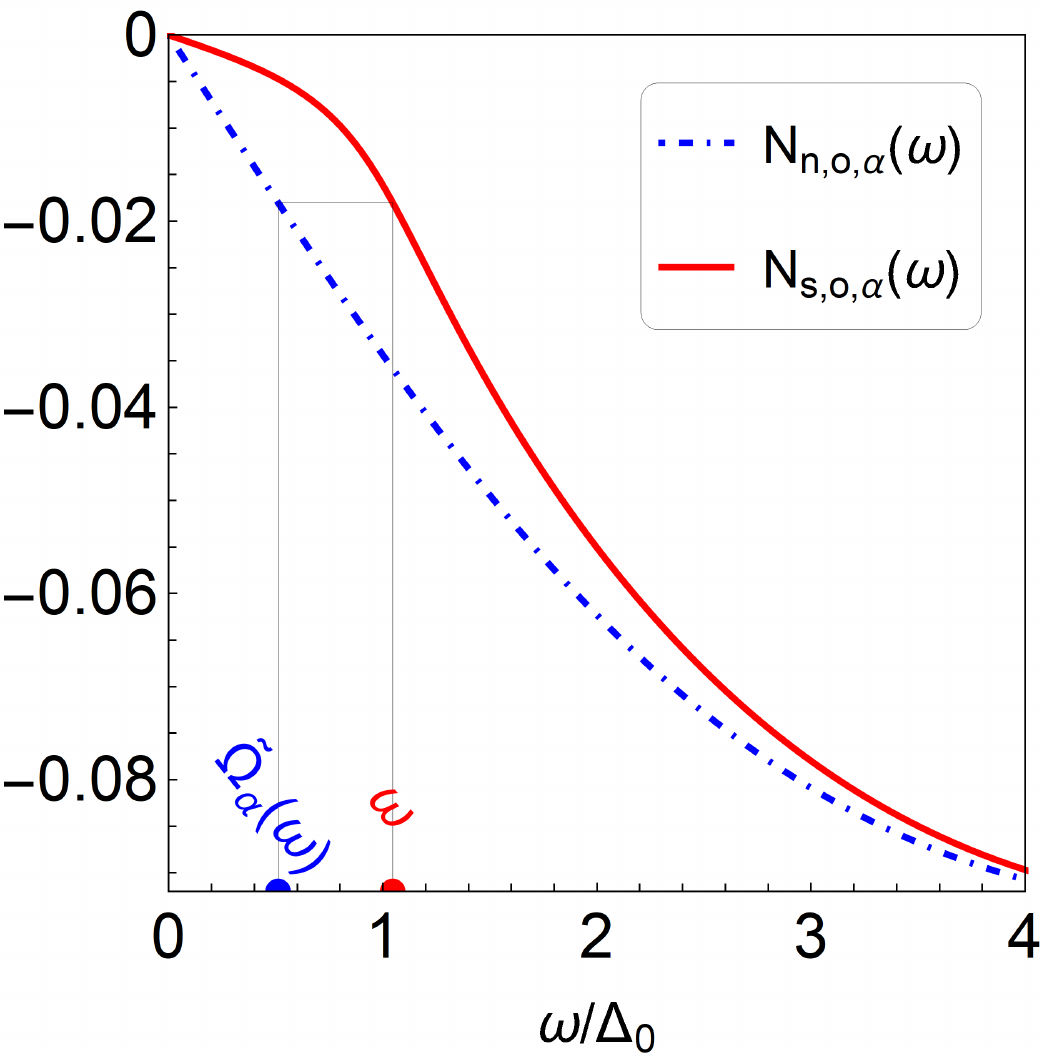}
\caption{Odd parts of the normal and superconducting densities of states, together with illustrated relation between the values $\omega$ and $\widetilde{\Omega}(\omega)$.}
\label{Fig:assym}
\end{figure}

\begin{figure}[h]
\includegraphics[width = 7.5 cm]{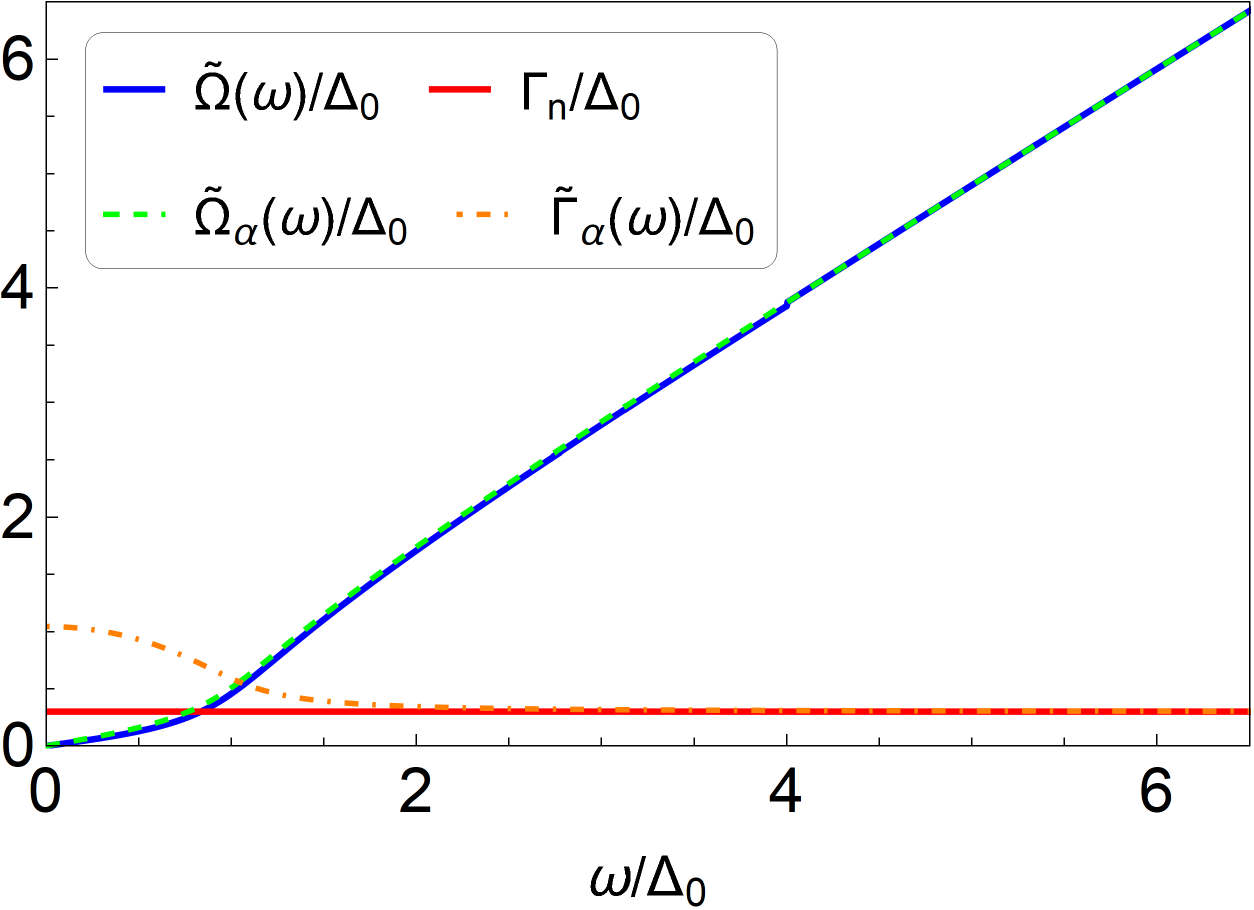}
\caption{Comparison of BM properties $\widetilde{\Omega}_{\alpha}$ and $\widetilde{\Gamma}_{\alpha}$ together with the ones assumed and determined so far by the solution of the Eq.~\eqref{eq:Nso_const_Gamma}.}
\label{Fig:tilde_Omega}
\end{figure}

First, let us focus on the Eq.~\eqref{eq:Nso_const_Gamma} relating the odd parts of  densities of states. We show the main idea of the solution in Fig.~\ref{Fig:assym}. Basically, we have to identify intervals on which are $N_{s,o}(\omega)$ and $N_{n,o}\big(\widetilde{\Omega}(\omega)\big)$ monotonous functions up to the point, where they actually merge. Next, inspecting the functions on the assumed interval, we determine the value of the $\widetilde{\Omega}(\omega)$ for which $N_{n,o}\big(\widetilde{\Omega}(\omega)\big) = N_{s,o}(\omega)$. 
The result of the described procedure considering the BM model is shown in Fig.~\ref{Fig:tilde_Omega}. We compare the assumed value of $\Gamma_n/\Delta_0$ and determined function $\widetilde{\Omega}(\omega)/\Delta_0$ together with the BM properties which result from the combination of Eq.~(\ref{eq:AuxialiaryFun}, \ref{eq:BM_Delta_Z}) and parameters specified in the caption of the Fig.~\ref{Fig:BM}.
From the definition of the auxiliary function $\widetilde{\Omega}(\omega)$ in Eq.~\eqref{eq:AuxialiaryFun}, we see, that once we are in the limit where the $\Delta(\omega)\rightarrow 0$, we can assume $\widetilde{\Omega}(\omega)\rightarrow\omega$.

\subsection{Solution for $\widetilde{n}(\omega)$}\label{Sec:tilde_n}

Here, we address the reconstruction of the $n(\omega)$, by solving the Eq.~\eqref{eq:Nse_const_Gamma}. In the beginning, purely for the purpose of having under control convergence of resulting integrals, let us rearrange Eq.~\eqref{eq:Nse_const_Gamma} into the form:
\begin{align}\label{eq:Nse_const_Gamma_Rearranged}
N_{s,e}(\omega) - N_{n,e}\big(\widetilde{\Omega}(\omega)\big) &= \big(n(\omega) - 1\big)N_{n,e}\big(\widetilde{\Omega}(\omega)\big)\\
&\qquad\qquad\qquad -\kappa(\omega)K_{n,o}\big(\widetilde{\Omega}(\omega)\big),\nonumber
\end{align}
Now, since $n(\omega)+i\kappa(\omega)$ is, up to an additive constant, the analytic function of $\omega$ in the upper-half plane of the complex space, we can expand $n(\omega)-1$ into a series of complex, analytic (in the upper-half plane of the complex space), rational eigenfunctions $\rho_n(x)$ of the Hilbert transformation, introduced in Ref.~\cite{Weideman95} and described in Appendix~\ref{Appendix:RatFun}:
\begin{align}
    n(\omega)-1 &= \sum_{k=-\infty}^{\infty} a_k \rho_k(\omega/\Theta).\label{eq:n}
\end{align}
The task is to find the values for the coefficients $a_k$.

After integrating both sides of Eq.~\eqref{eq:Nse_const_Gamma_Rearranged} multiplied by the complex conjugated eigenfunction $\rho^*_m(\omega)$ and using the orthogonal relation from Eq.~\eqref{eq:orthog}, we are left with the linear set of equations, defined as:
\begin{equation}\label{eq:cm}
b_m = \mathbb{M}_{ml}a_l,
\end{equation}
where, of course:
\begin{align}
    b_m &= \frac{1}{\pi}\int_{-\infty}^{\infty}d\omega \rho^*_m(\omega) \left[N_{s,e}(\omega) - N_{n,e}\big(\widetilde{\Omega}(\omega)\big)\right],\label{eq:Nne}\\
\mathbb{M}_{ml} &=\frac{1}{\pi} \int_{-\infty}^{\infty}d\omega \rho^*_m(\omega)\rho_l(\omega)\times\nonumber\\
&\qquad \left[N_{n,e}\big(\widetilde{\Omega}(\omega)\big) + i {\rm sgn}(l) K_{n,o}\big(\widetilde{\Omega}(\omega)\big)\right].\label{eq:M}
\end{align}
Now, we can inverse the matrix $\mathbb{M}$ introduced in the Eq.~\eqref{eq:cm} and express the coefficients $a_l$ as:
\begin{equation}\label{eq:dosfun_coef}
a_l = \mathbb{M}^{-1}_{lm} b_m.
\end{equation}
In the Appendix~\ref{Appendix:MandTech} we prove i) Matrix $\mathbb{M}$ is real. ii) In the special case, when $N_{n,e}(\omega)=const.$, matrix structure reduces to $\mathbb{M} = const. \mathbb{I}$, and $a_l = b_l/const.$, as it should.

Evaluating coefficients $b_m$ in the Eq.~\eqref{eq:Nne} and matrix elements $\mathbb{M}_{ml}$ from the Eq.~\eqref{eq:M} taking $-9\leq m,\,l\leq 9$ by using our BM and Eq.~\eqref{eq:dosfun_coef} we can easily get values of the coefficients $a_l$. Since the rational functions $\rho_k(x)$, used in the Eq.~\eqref{eq:n}, require characteristic energy scale $\Theta$, we use $\Theta = \Delta_0$.

\begin{figure}[htpb]
\includegraphics[width = 8.2 cm]{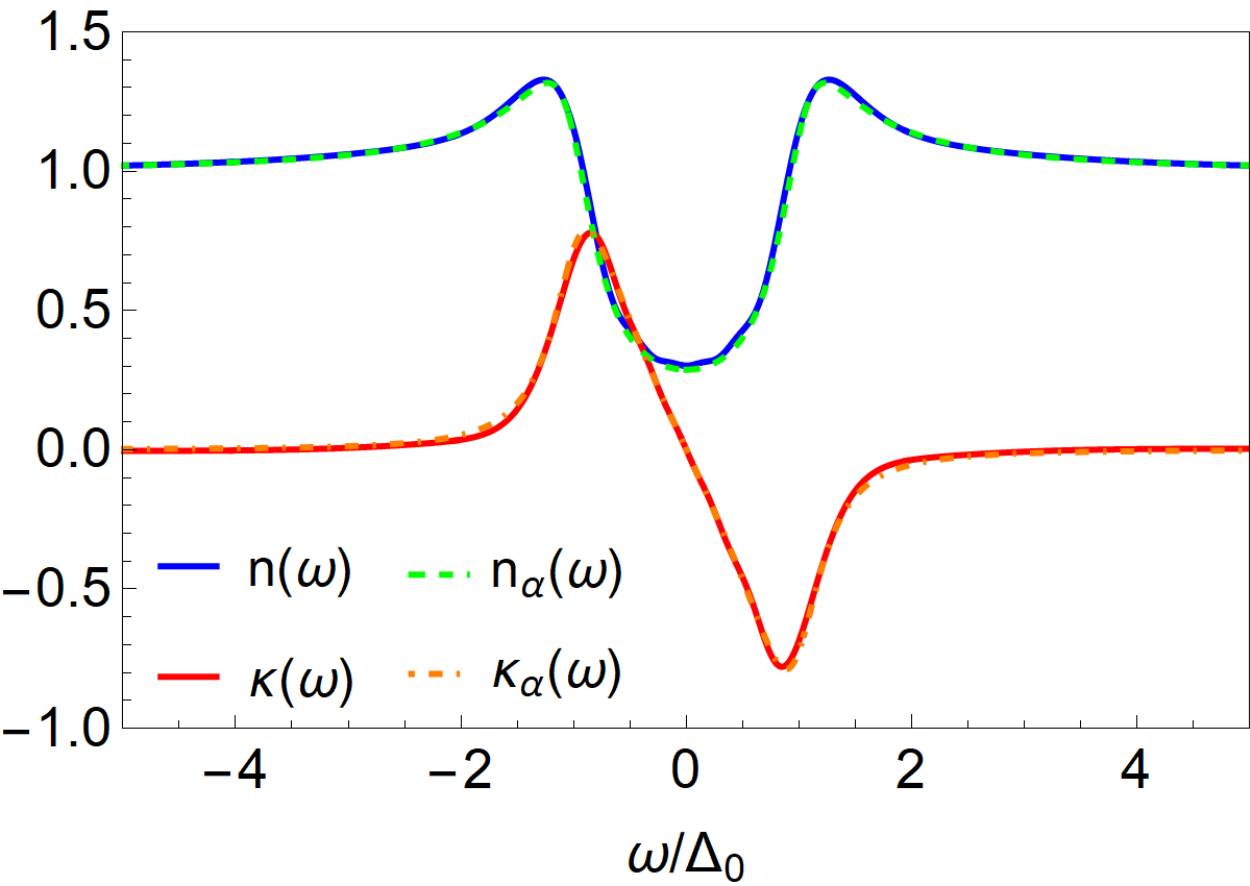}
\caption{Comparison of the calculated $n(\omega)$ and $\kappa(\omega)$ together with the BM, labeled by the index $\alpha$.}
\label{Fig:n_kappa}
\end{figure}

\begin{figure}[htpb]
\includegraphics[width = 7.5 cm]{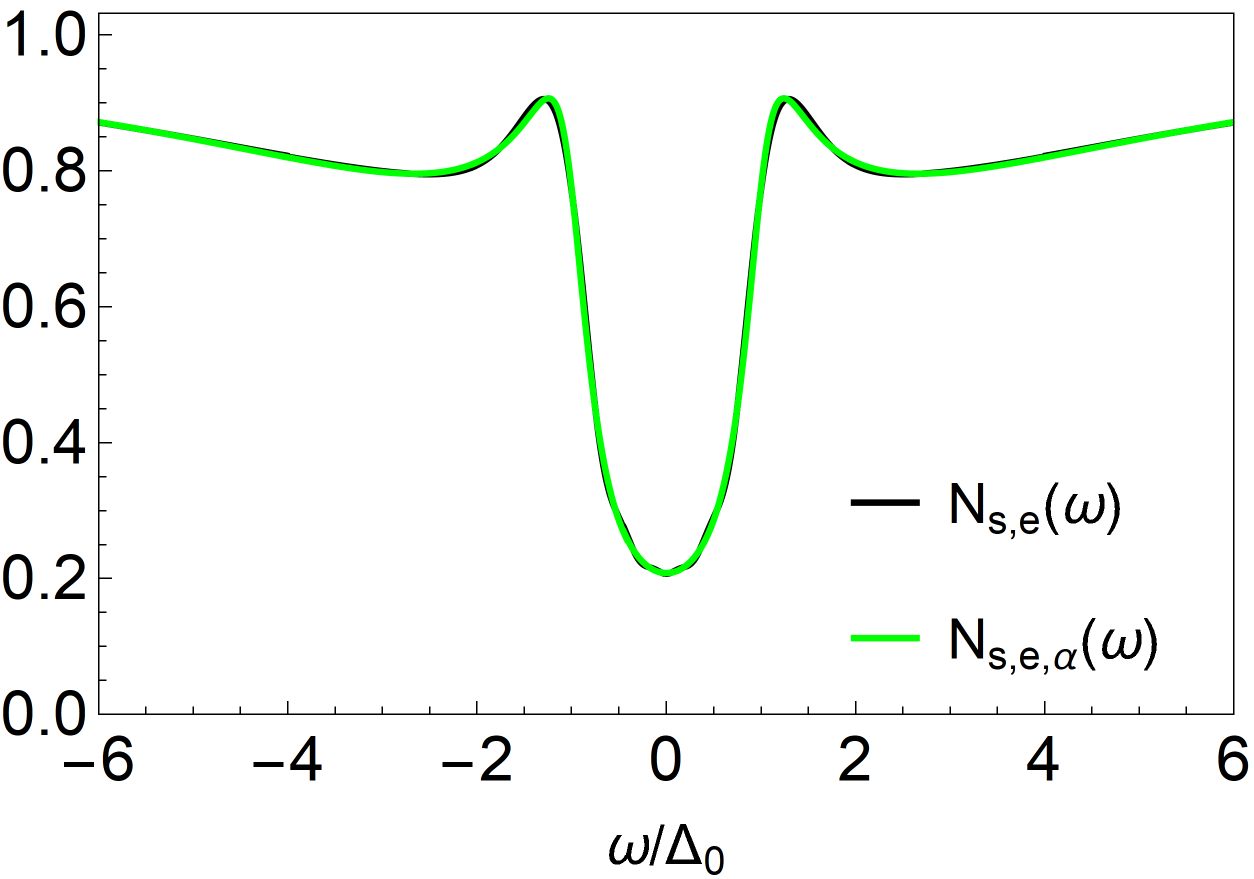}
\caption{Even parts of the BM superconducting density of states (green) and the one expanded to the set of eigenfunctions (see the text) $\rho_k(x)$ (black).}
\label{Fig:Ns_Nsmodel}
\end{figure}

Using Eq.~\eqref{eq:n} and Eq.~\eqref{eq:i_sgn} we can now express the fully reconstructed $\widetilde{n}(\omega)$. We compare our result marked as $\widetilde{n} = n+i\kappa$ with the BM Dynes-like $\widetilde{n}_{\alpha} = n_{\alpha}+i\kappa_{\alpha}$ in the Fig.~\ref{Fig:n_kappa}.
In Fig.~\ref{Fig:Ns_Nsmodel} we plot the full even part of the reconstructed density of state in the superconducting regime $N_s$, together with the even part of the original BM $N_{s,e,\alpha}$. Plotted comparison (or evaluated $N_s-N_{s,e,\alpha}$) can be used in order to fix a sufficient number of the used eigenfunctions $\rho_k(x)$ and energy scale $\Theta$. An unnecessarily large value of $\Theta$ causes the eigenfunctions $\rho_k(x)$ to drop to $0$ very slowly, which is a problem, since we integrate within the finite region in Eqs.~\eqref{eq:Nne} and \eqref{eq:M} (simulating the case of having the experimental data). On the other side, a very small value of $\Theta$ forces us to consider a large number of $\rho_k(x)$. The optimal value of $\Theta$ can minimize the difference of $|N_s-N_{s,e,\alpha}|$, together with a reasonable number of the considered eigenfunctions.

\subsection{Direct inversion for $\Delta(\omega)$}\label{Subsec:Direct_Inversion}

\begin{figure}[h!]
\includegraphics[width = 7.5 cm]{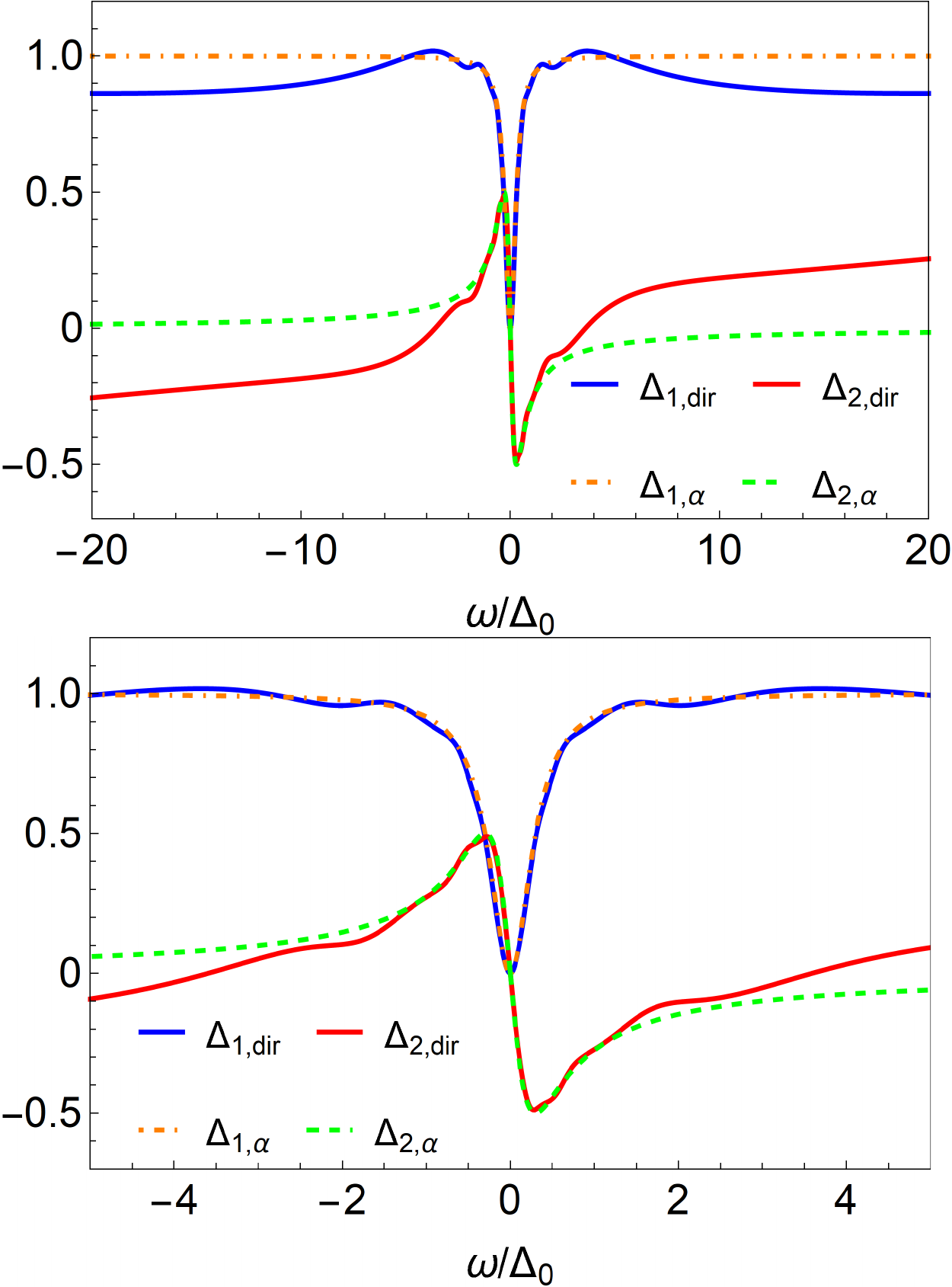}
\caption{Gap function $\Delta_{\textrm{dir}}(\omega)/\Delta_0$ (upper panel), together with its detail (lower panel), obtained by direct inversion and compared to the original gap function $\Delta_{\alpha}(\omega)/\Delta_0$, from the BM being the Dynes superconductor.}
\label{Fig:Delta_Direct}
\end{figure}

Within this section, we apply the $\Delta(\omega)$ extracting method originally proposed in Ref.~\cite{Galkin74}. We use it in the same manner as in Ref.~\cite{Kavicky22}:
\begin{equation}\label{eq:Delta_Direct}
    \Delta_{\textrm{dir}}(\omega) = \omega \sqrt{1-\widetilde{n}(\omega)^{-2}}.
\end{equation}
Obtained results describing real and imaginary part of the $\Delta_{\textrm{dir}}(\omega) = \Delta_{1, \textrm{dir}}(\omega) + i\Delta_{2, \textrm{dir}}(\omega)$ can be seen in the Fig.~\ref{Fig:Delta_Direct}.
Despite the simplicity of the direct inversion method, we can immediately witness its drawbacks in cases when $|\omega| \gtrsim |\Delta_0|$. As can be seen from the Eq.~\eqref{eq:comDOS}, or Eq.~\eqref{eq:Delta_Direct}, the accuracy of the determined $\Delta_{\textrm{dir}}(\omega)$ is suppressed, since this region, even for the relatively precisely reconstructed $\widetilde{n}(\omega)$ (see Fig.~\ref{Fig:n_kappa}), is dominated by the value of $\omega$. However, while we limit ourselves to the region below the gap, $|\omega| \lesssim |\Delta_0|$, we see that the extraction of $\Delta_{\textrm{dir}}(\omega)$ leads to a plausible result in comparison with the considered BM.

\subsection{Fitting for $\Delta(\omega)$}\label{Subsec:Fitting_for_Delta}

As we have seen in the Subsec:~\ref{Subsec:Direct_Inversion}, if we want to extract the gap function out of the already reconstructed $\widetilde{n}(\omega)$, we need a more accurate method. Therefore, motivated by the Ref.~\cite{Kavicky22} we use the verified analytic eigenfunctions from the Appendix~\ref{Appendix:RatFun} once again. Now, we expand the searched gap function $\Delta_{\textrm{fit}}(\omega)$ as:
\begin{equation}\label{eq:Delta_fit}
    \Delta_{\textrm{fit}}(\omega) = \Delta_0 + 2\sum_{k=0}^{N-1} c_k \rho_k\left(\frac{\omega}{\Lambda}\right),
\end{equation}
where we also need to specify: i) The number $N$, determining the number of $\rho_k(x)$ functions and so, how many coefficients $c_k$ we need to fix. ii) The energy scale $\Lambda$.

In order to fix $\Lambda$, we use an educated guess, based on the results from the previous section plotted in Fig.~\ref{Fig:Delta_Direct}. First, we limit ourselves to the region $|\omega|\ll|\Delta_0|$. Second, even without assuming any particular microscopic model, in this regime, we can simply suggest $\Delta(\omega) \approx \big(\omega/\lambda \big)^2-i \big(\omega/\lambda\big)$. Following fitting results to the value of $\lambda = 0.32 \Delta_0$ and reflecting the fact that this scale can be in principle valid only close to $\omega \sim 0$, we take this value into reasonable consideration and at the end, we fix $\Lambda = 0.2 \Delta_0$.

\begin{figure}[b]
\includegraphics[width = 8 cm]{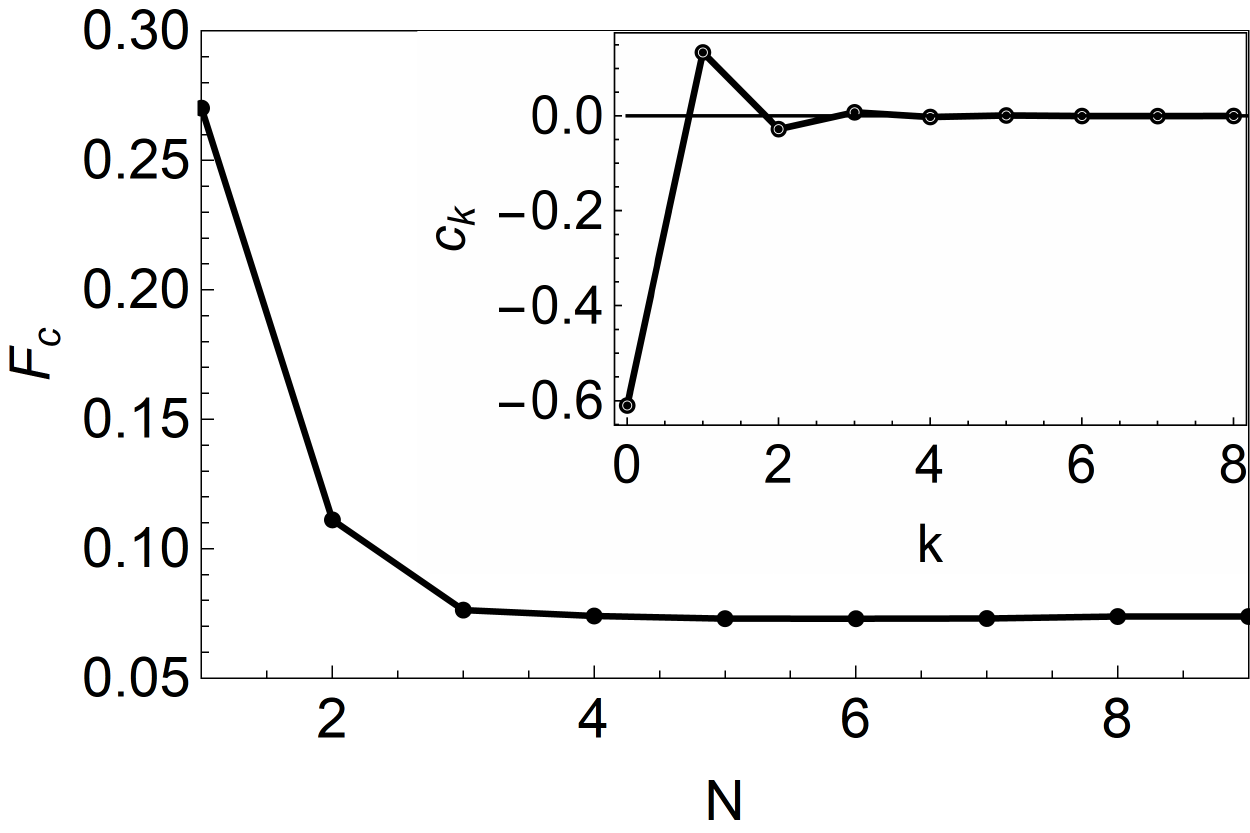}
\caption{Evolution of the cost function integral $F_c$ with the number of considered functions $\rho_k(x)$ and their optimal values for the coefficients $c_k$. The inset shows values of the coefficients $c_0,...,c_{8}$ minimizing $F_c$ defined in the Eq.~\eqref{eq:cost_fun}. The purpose of the black line is to guide the eye.}
\label{Fig:cost_fun}
\end{figure}

Next, to find values of the coefficients $c_k$ in the Eq.~\eqref{eq:Delta_fit} we define the following cost-function integral:
\begin{multline}\label{eq:cost_fun}
    F_c\big(\lbrace c_0,...,c_{N-1}\rbrace\big) \equiv \int_0^{\omega_0}d\omega \left| \frac{\widetilde{n}\big(\Delta_{\textrm{fit}}(\omega),\omega\big)}{\widetilde{n}(\omega)}-1\right|\\
    +\int_{\omega_1}^{\infty} d\omega\left|\frac{\Delta_{\textrm{fit}}(\omega)}{\Delta_{\infty}(\omega)}-1\right|,
\end{multline}
which consists of two contributions. The first one, dominating the cost-function value, uses $\widetilde{n}(\omega)$, which is the function determined by the expansion in Eq.~\eqref{eq:n} together, with the  coefficients $a_l$ determined from Eq.~\eqref{eq:dosfun_coef}. The function $\widetilde{n}\big(\Delta_{\textrm{fit}}(\omega),\omega\big)$ is Eq.~\eqref{eq:comDOS}, combined together with Eq.~\eqref{eq:Delta_fit}.\footnote{We remind the reader, that the known or assumed attributes used for constructing $F_c$ are coming from the so far reconstructed $\widetilde{n}(\omega)$ (not $\widetilde{n}_{\alpha}(\omega)$, describing BM).} The scale $\omega_0$ represents the natural boundary beyond which $\tilde{n}(\omega)\approx 1$. For our purposes, we assume $\omega_0 = 5\Delta_0$.

The second, order of magnitude smaller, term of Eq.~\eqref{eq:cost_fun} represents just the book-keeping property of the gap function approaching the BCS-like real-part-constant limit form $\Delta_{\infty}(\omega)\approx \Delta_0 + i \Delta_{2, \infty}(\omega)$ assuming $\omega\geq\omega_1\gg\Delta_0$. The imaginary part $\Delta_{2, \infty}(\omega)\propto 1/\omega$ preserves the analytic behavior maintained by the Kramers-Kronig relations. The full form of $\Delta_{\infty}(\omega)$ results from the sum rule derived and fully explained in detail in Appendix~\ref{Appendix:Gap_in_infty}:
\begin{eqnarray}\label{eq:Delta_infty_omega}
\Delta_{\infty}(\omega) \approx \Delta_0 + i\frac{2\Lambda}{\omega} \sum_{k=0}^{N-1}(-1)^k c_k.
\end{eqnarray}
In our case, we fix $\omega_1 = 10\Delta_0$, which justifies all the assumptions used in Appendix~\ref{Appendix:Gap_in_infty}.

We plot the results of the numerical minimization of the $F_c$ cost-function integral from Eq.~\eqref{eq:cost_fun} as a function of $N$ in Fig.~\ref{Fig:cost_fun}. As we expect, with the increasing number of considered coefficients $N$, the value of $F_c$ can be optimized to lower values. However, upon adding more optimization parameters in the form of the coefficients $c_k$, the parameter space grows as well and so does the required time for their optimization. In our case, it proves sufficient to explore $N \leq 9$.\footnote{It is natural to expect that one will need a similar number of eigenfunctions $\rho_k(x)$ to describe the gap function as was needed in order to reconstruct $\widetilde{n}(\omega)$ in Sec.~\ref{Sec:tilde_n}.}

\begin{figure}[h]
\includegraphics[width = 7.4 cm]{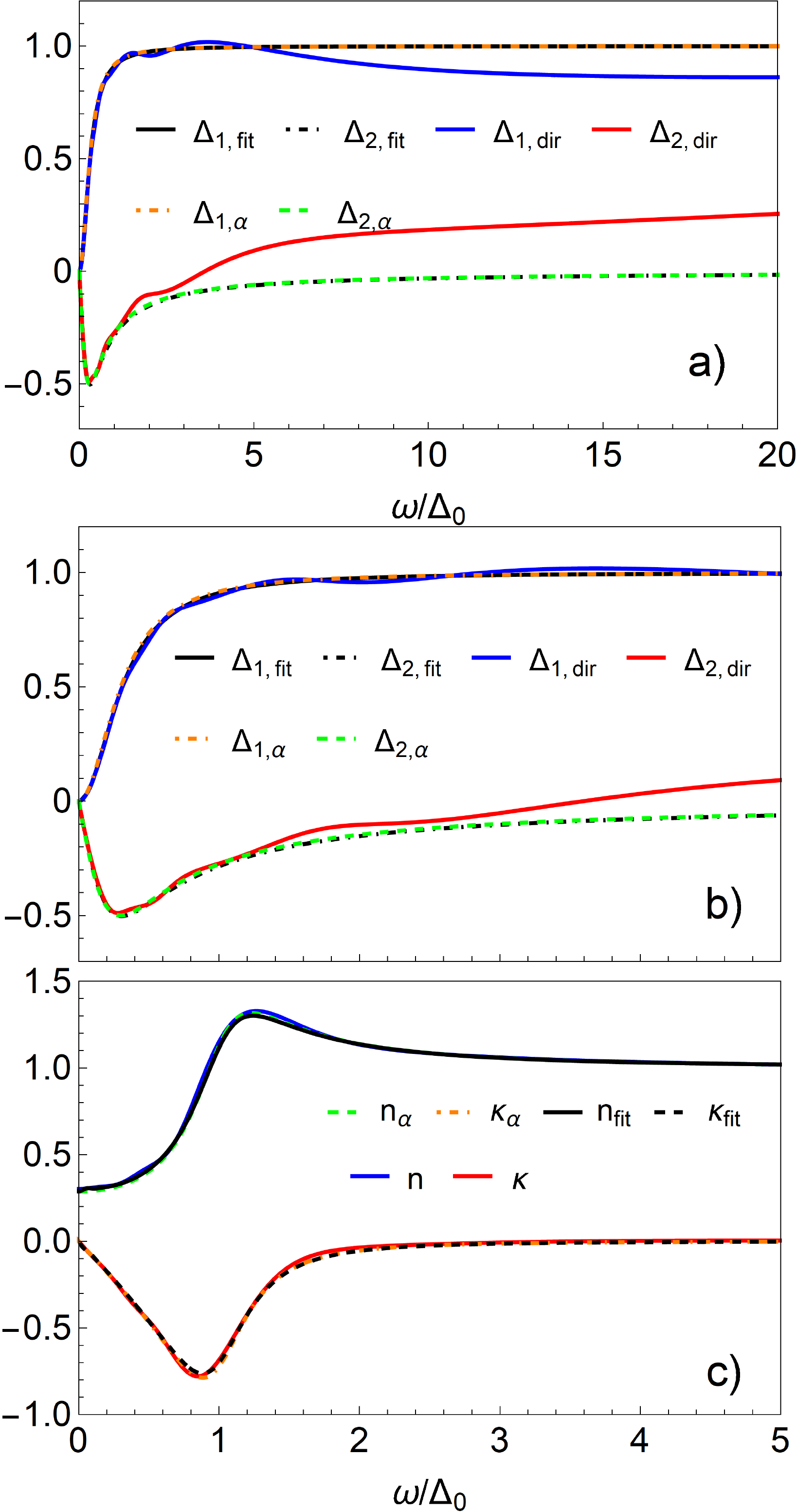}
\caption{a) Gap functions reconstructed by fitting $\Delta_{\textrm{fit}}(\omega)/\Delta_0$ (black and black-dot-dashed lines) compared with the BM Dynes gap function $\Delta_{\alpha}(\omega)/\Delta_0$ (orange-dot-dashed and green-dashed lines). For completeness, we show also the directly inverted gap function $\Delta_{\textrm{dir}}(\omega)/\Delta_0$ (blue and red). Panel b) captures details at lower energies $0\leq\omega/\Delta_0\leq5$. Control panel c) shows $\widetilde{n}(\omega)$ using gap functions from a) and b).}
\label{Fig:Final}
\end{figure}

We show the resulting optimized coefficients $c_k$ for $N=9$ in the inset of Fig.~\ref{Fig:cost_fun}. Notice, that the absolute value of the coefficients is going to zero with increasing $k$. We show results for the resulting gap function $\Delta_{\textrm{fit}}(\omega)$ compared with the BM model $\Delta_{\alpha}(\omega)$ and $\Delta_{\textrm{dir}}(\omega)$ (all in units of $\Delta_0$) in Fig.~\ref{Fig:Final}. As we can see, the fitting method works with a margin of error on the level of the thickness of the line for all considered energies. We consider this accuracy to be sufficient for our particular example. In Appendix~\ref{Appendix:MCBM} we elaborate on the reconstruction of the even more complicated model, including a more challenging form of the microscopic gap function and noise.

Coming back to the message of the inset in Fig.~\ref{Fig:cost_fun}, we can see that the most important coefficient $c_0$ is in absolute value close to the value $\Delta_0/2$, meanwhile, the rest is suppressed. Our results are much more easily understood once we realize, that the sought after $\Delta(\omega)$ coming from the superconducting part of our BM model defined by Eq.~\eqref{eq:BM_Delta_Z} can also be written as $\Delta_{\alpha}(\omega)/\Delta_0 = 1 - \rho_0\left(\omega/\Gamma\right)$. Once we compare it with our $\Delta_{fit.}(\omega)$ expansion defined in Eq.~\eqref{eq:Delta_fit}, we immediately notice that in the case of the perfectly reconstructed $\widetilde{n}(\omega)$ and properly estimated scale $\Lambda$, we should have just one nonzero series coefficient $a_0 = -\Delta_0/2$. Our numeric procedure confirms this idea nicely, once we fix the scale $\Lambda = \Gamma$. The values of the coefficients plotted in the inset of Fig.~\ref{Fig:cost_fun} are close to this behavior, since the estimated value of $\Lambda=0.2\Delta_0$ is close to our chosen value of the pair-breaking scattering rate $\Gamma = 0.3\Delta_0$. This sanity check is also one of the reasons why we considered the superconducting sector of the BM in the form of the DS in the first place.

\section{Conclusions}
To briefly summarize the main results, we conclude that our developed extraction method is shown to work accurately in cases with relatively simple behavior of the microscopic gap function. Results obtained in Appendix~\ref{Appendix:MCBM} suggest, that in cases with more effects on, and with different energy scales, as well as included noise, the extraction method is shown to have a decent qualitatively informative value. Separation of the superconducting dos function $n(\omega)$ and normal state density of states $N_n(\omega)$ works well for both, BM as well as MCBM. Computational time reduction in cases when we want to include a large number of basis functions within the fitting method represents one of the directions for further development and improvement of the introduced method.

\subsection*{Appllication to data}

Of course, the plan is to apply our method to the data from the experiment. Let us therefore summarize the main required properties of the appropriate data. In principle, we need the tunneling densities of states in the normal and superconductive state, which fulfill the following requirements: (i) They are measured on the sufficiently large energy intervals $|\omega|\lesssim 20|\Delta_0|$, with high resolution $\delta \omega \lll \Delta_0$, so we can accurately construct their Kramers-Kronig partners; and (ii) they have nonzero odd parts and low enough noise to identify these odd parts in order to exploit Eq.~\eqref{eq:Nso_const_Gamma}.

Published data, that are almost suitable for our method are found in Ref.~\cite{Carbillet20}. The low-temperature measurements of the tunneling conductance $dI/dV$ in the superconducting state could potentially be used as the input superconducting density of states. Published data have low noise, and very good resolution, and they are measured in the wide interval of $(-27\, mV,\, 27\, mV)$. Meanwhile, the value of the superconducting gap is roughly $1.2\,mV$. However, the biggest drawback, considering our purposes, is the lack of measurements of the low-temperature tunneling conductance in the normal state (e.g. above the transition driven by the external magnetic field). If those data were provided, we could proceed. Nevertheless, this example clearly shows that our procedure, in principle, is technically completely feasible.

\section*{Acknowledgements}

I acknowledge fruitful discussions with Dušan Kavický and Richard Hlubina. I am also grateful for the financial support from the Slovak Research and Development Agency under Contract No. APVV-19-0371, by the agency VEGA under Contract No. 1/0640/20 and by the European Union's Horizon 2020 research and innovation programme under the Marie Sk{\l}odowska-Curie Grant Agreement No.~945478.

\clearpage

\begin{appendix}

\section{Rational functions}\label{Appendix:RatFun}

In this appendix, we remind the reader of the basic properties of rational, analytic (in the upper half-plane of the complex space), orthogonal basis functions from Ref.~\cite{Weideman95}, defined in the form:
\begin{equation*}
\rho_n(x)=\frac{(1+ix)^n}{(1-ix)^{n+1}}, \quad n=0,\pm 1,\pm2, \ldots,
\end{equation*}
together with their orthogonality relation,\footnote{It is good to remember, that once we use this mathematical equation in physics, where we are confronted with dimensional quantities, we are left with:
$$
\frac{1}{\pi\Lambda}\int_{-\infty}^\infty d\omega \rho_n^\ast\left(\frac{\omega}{\Lambda}\right)\rho_m\left(\frac{\omega}{\Lambda}\right) = \delta_{nm}.
$$
where $\Lambda$ is a suitable energy constant.
}:
\begin{equation}\label{eq:orthog}
\frac{1}{\pi}\int_{-\infty}^\infty dx \rho_n^\ast(x)\rho_m(x)= \delta_{nm}.
\end{equation}
\begin{figure}[htpb]
\includegraphics[width = 7.3 cm]{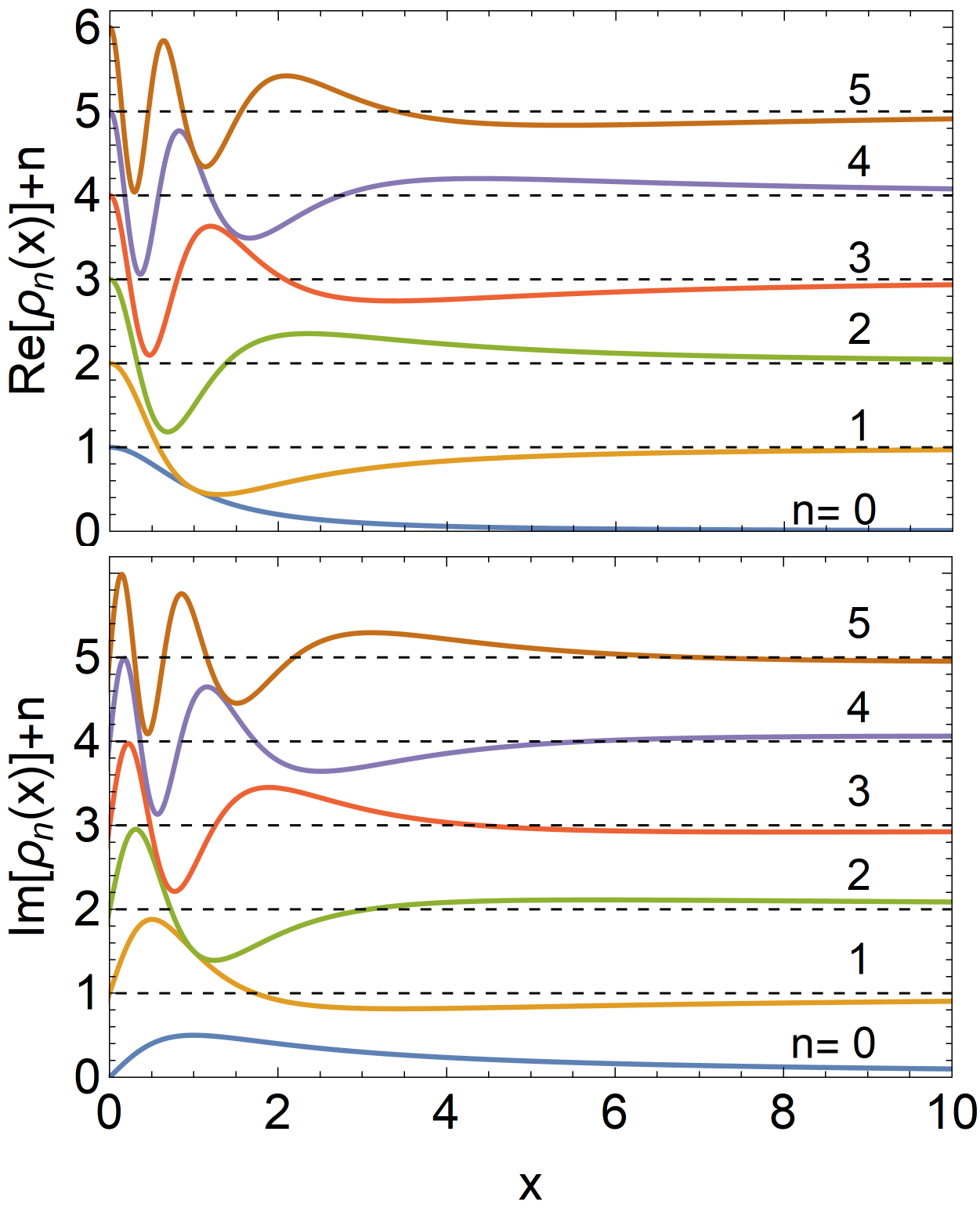}
\caption{Real (upper panel) and imaginary (lower panel) parts of the basis functions $\rho_n(x)$ for $n=0,\ldots,5$. Note that, with increasing $n$, the period of oscillations at small $x$ increases and, at the same time, the largest nodes $x_n$ of both components grow \cite{Kavicky22}. The overall shape of the basis functions explains why a more lengthily expanded function will require a larger number of eigenfunctions, in order to reconstruct it accurately.}
\label{fig:basis_functions}
\end{figure}

Equation \eqref{eq:orthog} implies that the coefficients in the expansion of the function $f(x)=\sum_n a_n\rho_n(x)$ can be calculated as: 
\begin{equation*}
a_n=\frac{1}{\pi}\int_{-\infty}^\infty dx \rho_n^\ast(x) f(x).
\end{equation*}
The first few functions $\rho_n(x)$ are shown in Fig.~\ref{fig:basis_functions}.

The crucial point to observe is that $\rho_n(x)$ are eigenfunctions of the Hilbert transform \cite{Weideman95}, i.e. they satisfy:
\begin{equation}\label{eq:i_sgn}
\frac{1}{\pi}P\int_{-\infty}^\infty\frac{dx \rho_n(x)}{x-y}=i{\rm sgn}(n) \rho_n(y),
\end{equation}
where $P\int dx$ denotes the principal value integration.

\subsection*{Expansion of the constant function}

In order to demonstrate some of the properties of the $\rho_n(x)$, let us expand the simplest nonzero case $f(x)=1$. We can examine:
\begin{align}\label{eq:integr}
    \frac{1}{\pi}\int_{-\infty}^{\infty}dx \rho_n(x) &= \frac{1}{\pi}\int_{-\infty}^{\infty}dx \rho^*_n(x),
    \\ &=
    \begin{cases}
    (-1)^{n}, \mathrm{for}\, n\in \lbrace 0,1, ...,\infty\rbrace,\\
    \\
    (-1)^{n+1}, \mathrm{for}\, n\in \lbrace -\infty,...,-2,-1\rbrace.
    \end{cases}\label{eq:int_rho}
\end{align}
The proof (by using the induction method) assumes branch $k\in \lbrace 0,1, ...,\infty\rbrace$:
\begin{proof}
a) Case $k=0$, obviously:
\begin{equation*}
    \frac{1}{\pi}\int_{-\infty}^{\infty}dx \rho_0(x) = \frac{1}{\pi}\int_{-\infty}^{\infty}\frac{dx}{1-ix} = 1.
\end{equation*}
b) Proof for $n = k + 1$ assuming \eqref{eq:integr} is true for $n=k$, where $k\in \lbrace 0,1, ...,\infty\rbrace$:
\begin{align*}
    \frac{1}{\pi}\int_{-\infty}^{\infty}dx \rho_{k+1}(x) &= \frac{1}{\pi}\int_{-\infty}^{\infty}dx \rho_{k}(x)\frac{1+ix}{1-ix},\\
                                                        &= \frac{1}{\pi}\int_{-\infty}^{\infty}dx \rho_{k}(x)\left(\frac{2}{1-ix}-1\right),\\
                                                        &= \frac{2}{\pi}\int_{-\infty}^{\infty}dx \rho_{k}(x)\rho_0(x) - (-1)^{k},\\
                                                        &= \frac{2}{\pi}\int_{-\infty}^{\infty}dx \rho_{k}(x)\rho_{-1}^\ast(x) + (-1)^{k+1},\\
                                                        &= (-1)^{k+1}.\quad
\end{align*}
In the fourth line, we used symmetry relation $\rho_{l}(x) = \rho_{-l-1}^\ast(x)$, and in the fifth the orthogonality relation \eqref{eq:orthog}.
\end{proof}

Now, we focus on the proof of the second branch of the Eq.~\eqref{eq:int_rho}, which represents cases where $k\in \lbrace -\infty,...,-2,-1\rbrace$: 
\begin{proof}
At first, let us rewrite Eq.~\eqref{eq:integr}, which holds true generally, for all $k\in\mathbb{Z}$ in the form, which is the most illustrative for our purpose:
\begin{align*}
    \frac{1}{\pi}\int_{-\infty}^{\infty}dx \rho_{-|k|}(x) &= \frac{1}{\pi}\int_{-\infty}^{\infty}dx \rho^*_{-|k|}(x),\\
                                                        &= \frac{1}{\pi}\int_{-\infty}^{\infty}dx \rho_{|k|-1}(x),\\
                                                        &= (-1)^{|k|-1},\\
                                                        &= (-1)^{-k-1},\\
                                                        &= (-1)^{k+1},
\end{align*}
where, in the second line we used the symmetry $\rho_{-|l|}^{\ast}(x) = \rho_{|l|-1}(x)$. In the third line, we realized that the index $|k|-1\in \lbrace 0,1,...,\infty\rbrace$, so we can use the already proven result for the upper branch of the Eq.~\eqref{eq:int_rho}. In the fourth line, we realized that assuming $k<0$, $|k|=-k$. The fifth line is just trivial algebra in order to simplify the result and finish the proof of the lower branch of Eq.~\eqref{eq:int_rho}.
\end{proof}

The proven result is shown to be useful also for further purposes explained in the Sec.~\ref{Subsec:Fitting_for_Delta} and Appendix~\ref{Appendix:Gap_in_infty}. In this appendix, we just combine Eq.~\eqref{eq:Delta_fit} together with Eq.~\eqref{eq:int_rho}, so we are left with:
\begin{equation}\label{eq:c_k_int}
    \int_{-\infty}^{\infty}\frac{d\omega}{2\pi} \left[\Delta_{\textrm{fit}}(\omega)-\Delta_0\right] = \Lambda\sum_{k = 0}^{N-1} (-1)^{k} c_k.
\end{equation}

Expansion of the constant function $f(x)=1$, resulting in the Eq.~\eqref{eq:int_rho}, also shows that in principle we need to use an infinite amount of expansion coefficients. To avoid this behavior in our presented framework, we assume that the behavior of the superconducting properties naturally approaches those from the normal state, considering energy scales much larger than the superconducting gap. Therefore, we always expand the difference between the superconducting and normal state, which diminishes to zero for $|\omega|\ggg \Delta_0$.

\section{Properties of the matrix $\mathbb{M}$ and other technical details}\label{Appendix:MandTech}

In this appendix, we comment on a few properties of the matrix $\mathbb{M}$ introduced in Eq.~\eqref{eq:M} and technical details of the recipe described in Sec.~\ref{Sec:Solution}: 
\begin{itemize}
    \item[(i)] Matrix $\mathbb{M}_{ml}$ is real.
    \begin{proof}
    The structure of the integral is simply:
    \begin{equation}\label{eq:I}
        I = \int_{-\infty}^{\infty}d\omega\prod_{k=1}^{N}\big(e_k(\omega)+i o_k(\omega)\big),
    \end{equation}
    where $e(-\omega)=e(\omega)$, $o(-\omega)=-o(\omega)$ are real even, or odd functions and $N=3$ in our specific case. After the transformation $\omega \rightarrow -\omega$ inside $I$, we can immediately notice $I=I^*$ for any considered $N$, since for the imaginary part we integrate an odd function on the even interval. This means that $I$ in Eq.~\eqref{eq:I} is real.
    \end{proof}
    \item[(ii)] 
    If $N_{n,e}(\omega)=const.$, then: $\mathbb{M}_{ml} = const. \delta_{ml}$.
    \begin{proof}
    All we need to do is to realize that the square bracket in the Eq.~\eqref{eq:M} will be $const.$, since the Kramers-Kronig partner to a $const.$ is $0$. Next, we apply the orthogonality relation Eq.~\eqref{eq:orthog} and the coefficients $a_n = b_n/const.$.
    \end{proof}
    \item[(iii)] Also, in the numerical calculation,  infinities in \eqref{eq:n}, \eqref{eq:Nne}, \eqref{eq:M} will be replaced by finite (but sufficiently large) numbers.
\end{itemize}

\section{$\Delta_{\infty}(\omega)$ --- $\Delta_{\textrm{fit}}(\omega)$ assuming $\omega\gg\Delta_0$}\label{Appendix:Gap_in_infty}
We start with the full Kramers-Kronig equation for the imaginary part of the gap function:
\begin{equation}\label{eq:Delta_2}
    \Delta_2(\omega) = {\rm P}\int_{-\infty}^{\infty} \frac{d\omega'}{\pi}\frac{\Delta_1(\omega')-\Delta_0}{\omega-\omega'}.\\
\end{equation}
Now, assuming very large $\omega\gg\Delta_0$ representing the energy scale on which we can assume $\Delta_1(\omega)\approx \Delta_0$ and that the $\omega'$ term in the numerator of Eq.~\eqref{eq:Delta_2} is negligible, we can write:
\begin{equation}
    \Delta_{\infty,2}(\omega) \approx \int_{-\infty}^{\infty} \frac{d \omega'}{\pi \omega} \big(\Delta_1(\omega')-\Delta_0\big)\\
    = \frac{2\Lambda}{\omega} \sum_{k=0}^{N-1}(-1)^k c_k,
\end{equation}
where we used Eq.~\eqref{eq:c_k_int} in the last step in order to derive Eq.~\eqref{eq:Delta_infty_omega} in the main text.

\section{More Complicated Benchmark Model}\label{Appendix:MCBM}

In this appendix, we want to demonstrate, that the method introduced and explained in the main text can reconstruct an even more complicated model than the BM from the main text. In principle, the model that we show here is based on the most complicated theory model, reconstructed in Ref.~\cite{Kavicky22}. However, in order to go beyond its capabilities as well as fulfill the requirements of applicability of the approach presented in this manuscript, we add a non-zero odd part to the normal state behavior and we also add a completely random noise on top of the normal and also on the top of the superconducting state. The level of noise is kept at the level of $5\%$ of the maximal amplitude of the odd part of the normal state density of states.

\subsection*{Definition}

For further purposes, we refer to our more complicated model just as the more complicated benchmark model (MCBM). We define the superconducting part of the MCBM by the gap function $\Delta_{\beta}(\omega)$ (originally used in Ref.~\cite{Kavicky22}) in the form: 
\begin{align}
\Delta_{\beta}(\omega)&=\Delta_\infty + (\Delta_0-\Delta_\infty)F(\omega)
-\frac{i\Gamma \Delta_0}{\omega+i\Gamma},
\label{eq:model_MCBM_delta}
\\
F(\omega)&=\frac{i}{\pi}\left[\Psi\left(\frac{1}{2}+\frac{\omega+\omega_{*}}{2\pi i\eta}\right)-\Psi\left(\frac{1}{2}+\frac{\omega-\omega_{*}}{2\pi i\eta}\right)\right],
\nonumber\\
Z_{\beta}(\omega) &= \left(1 + \frac{i\Gamma}{\omega}\right).\nonumber
\end{align}
The model of the superconducting gap function $\Delta_{\beta}(\omega)$ aims at describing different effects of the superconducting behavior. Firstly, the gap value reduction is represented by the parameter $\Delta_0$ at the energy scale $\omega_{*}$, whereas the scale of the change is defined by the parameter $\eta$. Secondly, we also consider the effects of the Dynes-like pair-breaking scattering rate at a much smaller energy scale (in comparison with the value of the gap) represented by the term including parameter $\Gamma$.
\begin{figure}[htpb]
\includegraphics[width = 7.5 cm]{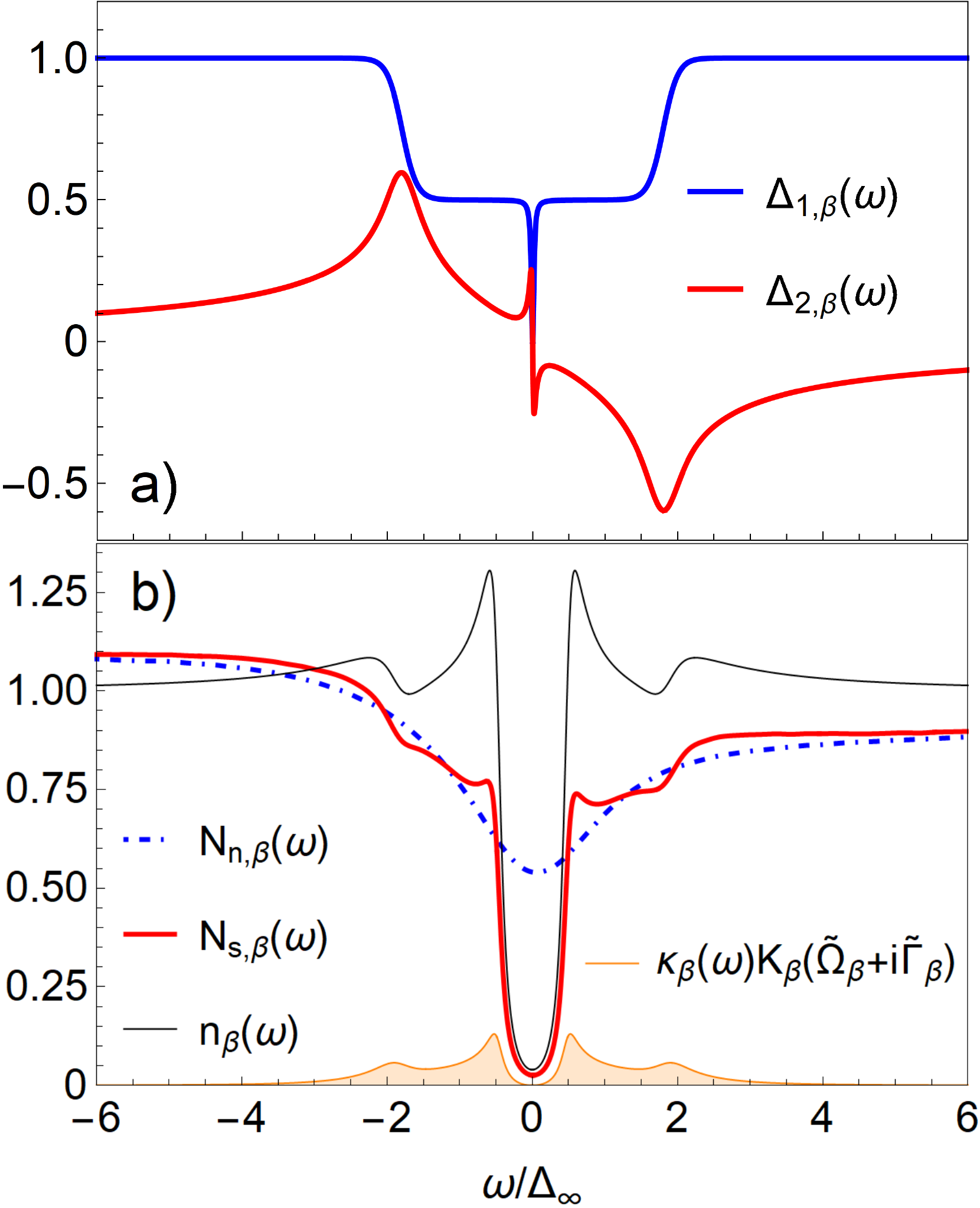}
\caption{
Subfigure a) presents the gap function of the MCBM model defined by the Eq.~\eqref{eq:model_MCBM_delta} normed on the value of the $\Delta_{\infty}$. We use parameter values $\Gamma/\Delta_{\infty} = 0.02$, $\Delta_0/\Delta_{\infty} = 0.5$, $\omega_*/\Delta_{\infty} = 1.8$, and $\eta/\Delta_{\infty} = 0.1$. In subfigure b) we evaluate the full MCBM defined by the Eq.~\eqref{eq:Nseo}. The normal state density of states $N_{n,\beta}(\omega)$, represented by the dot-dashed blue line, is defined by the Eq.~\eqref{eq:BM_N_0} using $\alpha_e = -0.466$, $\Gamma_e/\Delta_{\infty} = 1.2$, $\alpha_o = 0.2$ and $\Gamma_o/\Delta_{\infty} = 5$.
}
\label{Fig:MCBM}
\end{figure}

As in the main text, Fig.~\ref{Fig:MCBM} a) represents the gap function $\Delta_{\beta}(\omega)=\Delta_{1,\beta}(\omega)+i\Delta_{2,\beta}(\omega)$, which we want to reconstruct using data from the related densities of states.

To model the normal state, we use Eq.~\eqref{eq:BM_N_0} once again, but this time the normal state minimum in the density of states is significantly thinner, and deeper in comparison with the one presented in Fig.~\ref{Fig:BM} b) of the BM model. We present the evaluation of the MCBM properties in the normal and superconducting densities of states in Fig.~\ref{Fig:MCBM} b).

The effect of the Altshuler-Aronov minimum from the normal state is clearly visible if we compare the black line representing the superconducting dos function $n_{\beta}(\omega)$, with the red curve representing the actual density of states $N_{s,\beta}(\omega)$, with the minimum in the normal state. Notice the influence of the deeper minimum on the superconducting properties, which are: i) Relatively small impact inside the region of the smeared gap itself. ii) Significant influence on the magnitude of the coherence peaks. iii) Deformed signature from the reduction of the gap function at the scale $\omega_*$. MCBM also represents an example, which is the non-negligible effect of the second term of $N_{s,e}(\omega)$ from Eq.~\eqref{eq:Nseo} of the main text. We plot and highlight its absolute value with the orange curve in Fig.~\ref{Fig:MCBM} b).

\begin{figure}[t]
\includegraphics[width = 5.5 cm]{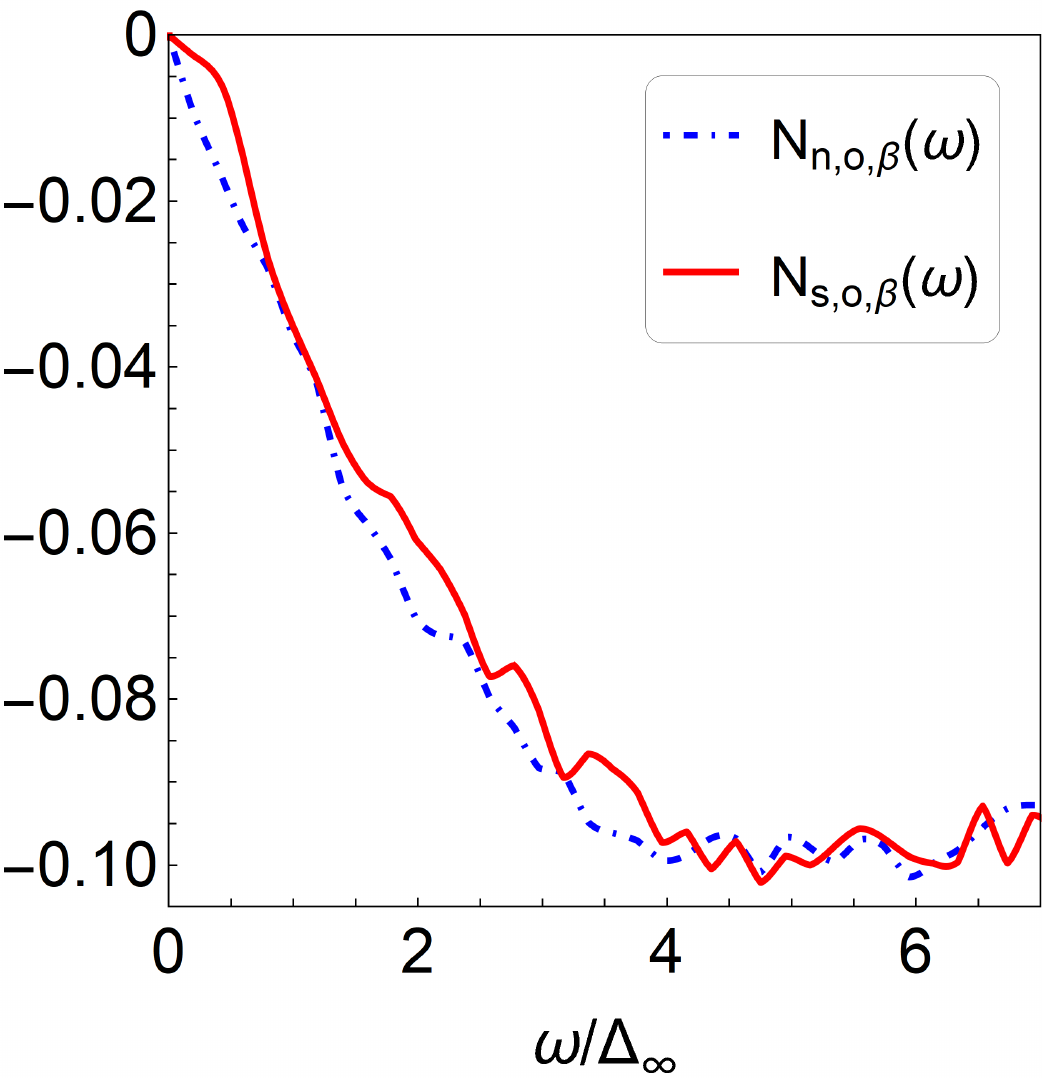}
\caption{Odd parts of the normal $N_{n,o,\beta}(\omega)$ and superconducting $N_{s,o,\beta}(\omega)$ densities of states for the MCBM model.}
\label{Fig:assym_MCBM}
\end{figure}

The odd parts of the densities of states are presented in Fig.~\ref{Fig:assym_MCBM}. It has been created in the same manner as the one using the BM model and presented in Fig.~\ref{Fig:assym} of the main text. In Fig.~\ref{Fig:assym_MCBM}, we can also notice the level of the included random noise.

\subsection*{Solution}

Having fully defined the properties of the MCBM model, we can take the data of the resulting densities of states and we can proceed to the extraction of the superconducting properties. 

\begin{figure}[htpb]
\includegraphics[width = 7.1 cm]{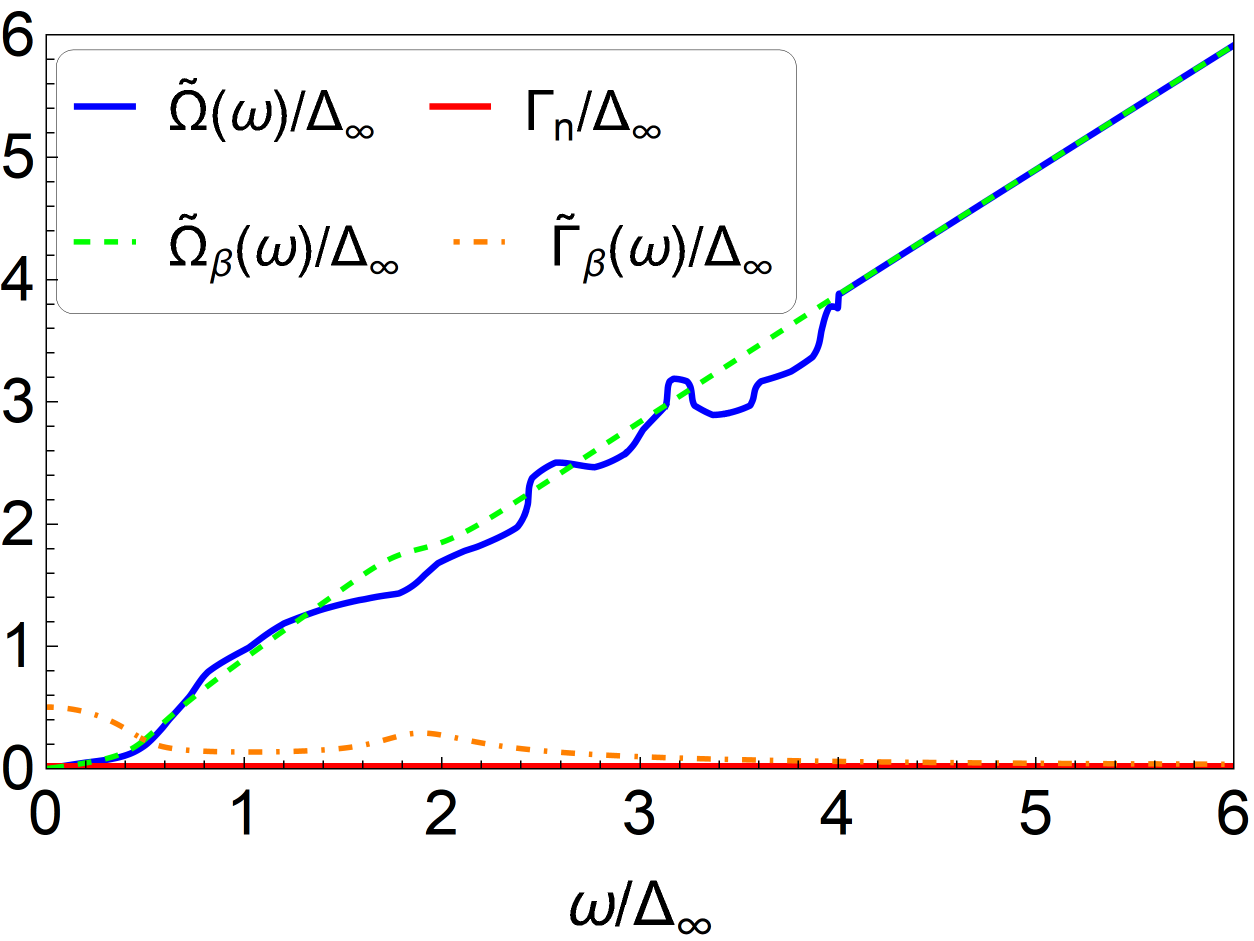}
\caption{Comparison of MCBM properties together with the ones we assumed and determined so far by solving the Eq.~\eqref{eq:Nso_const_Gamma}.}
\label{Fig:tilde_Omega_MCBM}
\end{figure}

{\it Solution for $\widetilde{\Omega}(\omega)$.} First, following the procedure described in Sec.~\ref{SubSec:Sol_Tilde_Omega}, we proceed to the reconstruction of $\widetilde{\Omega}(\omega)$. We show the result in Fig.~\ref{Fig:tilde_Omega_MCBM}, where we can also recognize the effect of random noise. Let us note that we assume $\widetilde{\Omega}(\omega) = \omega$, considering $\omega\geq 4\Delta_{\infty}$.
Focusing on the results of the reconstructed $\widetilde{\Omega}(\omega)$ (blue curve) with the actual MCBM model (green dashed curve), we are overall satisfied with the result so far.

\begin{figure}[b]
\includegraphics[width = 7.5 cm]{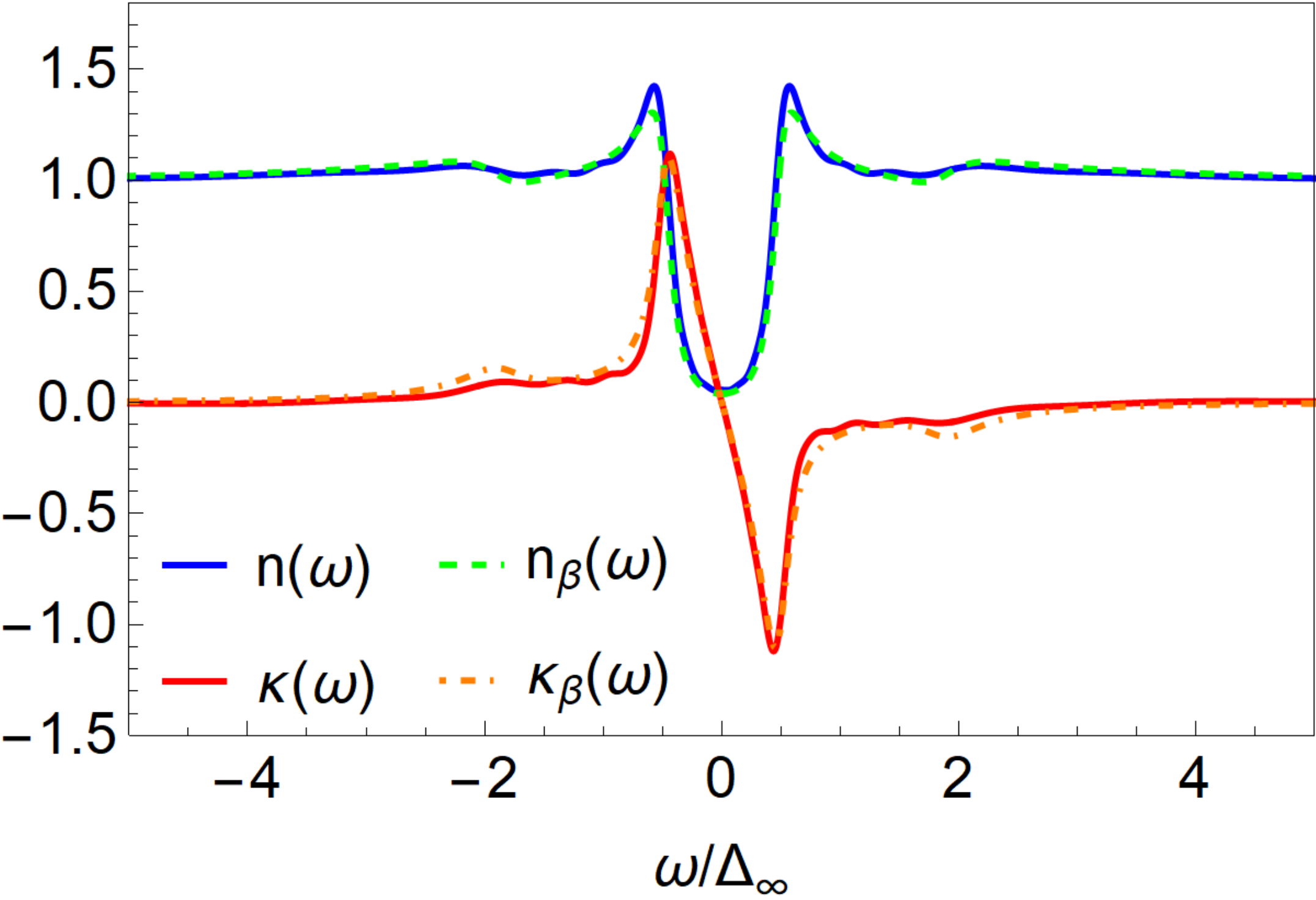}
\caption{Comparison of the calculated $n(\omega)$ and $\kappa(\omega)$ together with the MCBM.}
\label{Fig:n_kappa_MCBM}
\end{figure}

{\it Solution for $\widetilde{n}(\omega)$.} Next, we continue with the reconstruction of $\widetilde{n}(\omega)$ based on the Sec.~\ref{Sec:tilde_n}. Since the whole recipe is already fully described in that section, together with the referenced appendixes, we can skip directly to the results, presented in Fig.~\ref{Fig:n_kappa_MCBM}.

As we can notice, our method can nicely and qualitatively reconstruct all of the incorporated properties of the superconducting state on the level of $n(\omega)$ and $\kappa(\omega)$, despite the deep suppression,  (rather harsh) assumption $\widetilde{\Gamma}(\omega) \approx \Gamma_n = \Gamma$, or the included noise. 

Let us note, that this result is important on its own since $n(\omega)$ describes the properties of the superconductor with the constant density of state in the normal state. The properties of the $\kappa(\omega)$ are guaranteed by being the Kramers-Kronig partner of $n(\omega)$ since we compound the analytic $\widetilde{n}(\omega)$ from the analytic basis functions $\rho_n(\omega)$.

\begin{figure}[htpb]
\includegraphics[width = 7.3 cm]{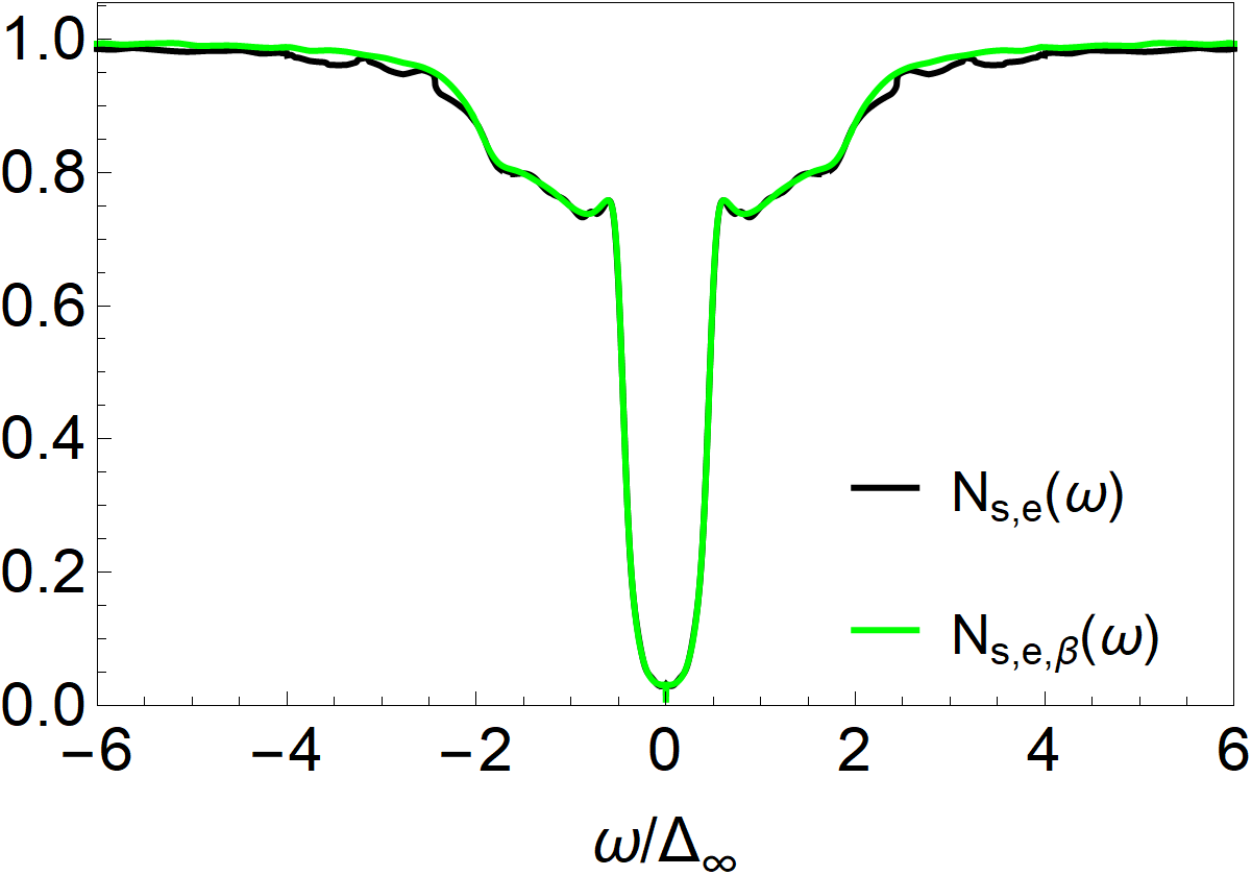}
\caption{Even parts of the MCBM superconducting density of states (green) and the reconstructed one (black).}
\label{Fig:Ns_Nsmodel_MCBM}
\end{figure}

We also show a comparison of the even part of the superconducting part of the original model $N_{s,e,\beta}(\omega)$, together with $N_{s,e}(\omega)$, being expanded to the basis functions $\rho_k(x)$, in Fig.~\ref{Fig:Ns_Nsmodel_MCBM}. Despite all of the mentioned complications, the reconstruction of the superconducting state has turned out quite well until now. Note, that we assume $20$ basis functions $\rho_k(\omega)$.

\begin{figure}[htpb]
\includegraphics[width = 7.1 cm]{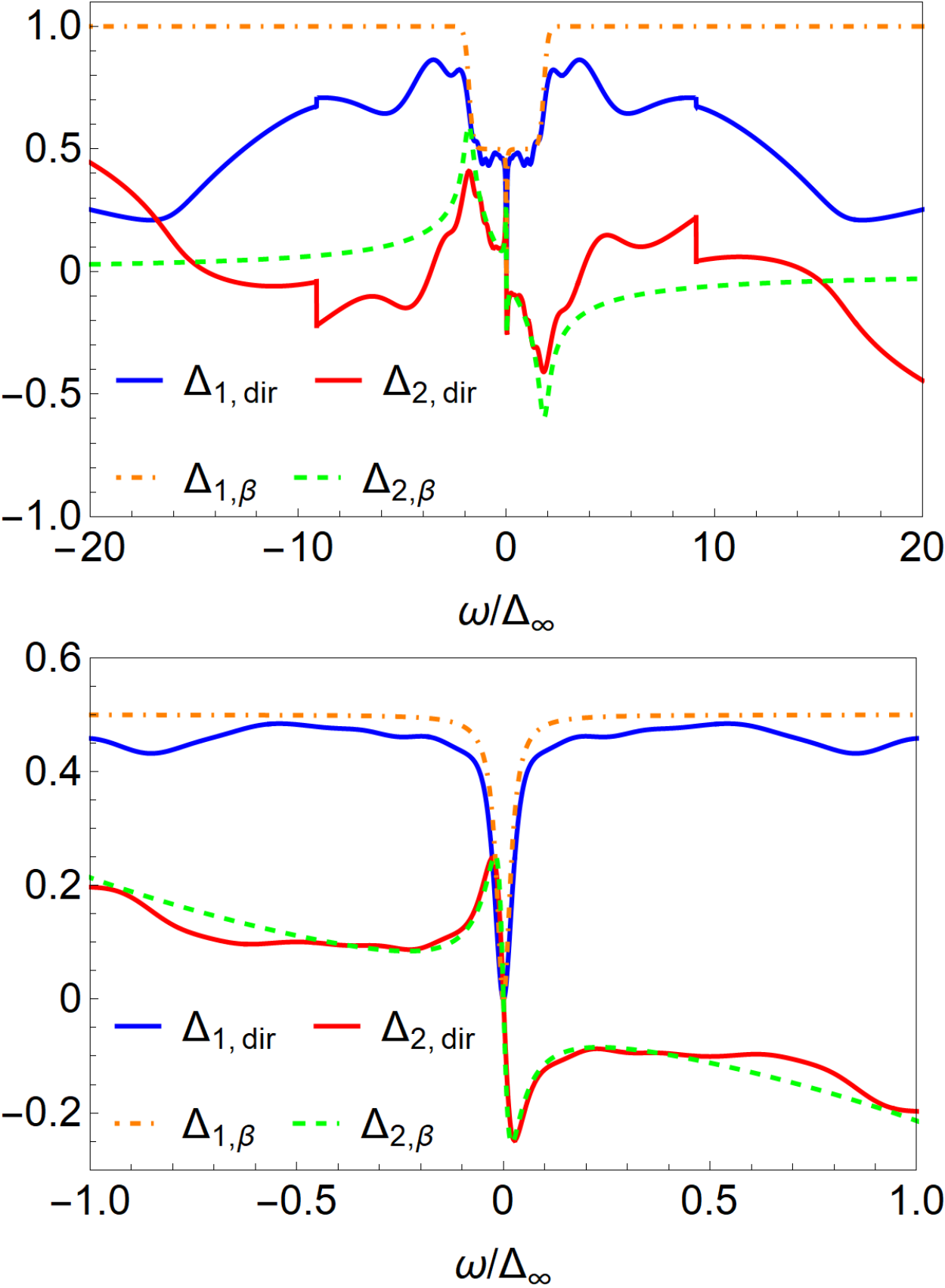}
\caption{Gap function $\Delta_{\textrm{dir}}(\omega)/\Delta_{\infty}$ (upper panel) together with its detail (lower panel) obtained by the direct inversion and compared to the gap function $\Delta_{\beta}(\omega)/\Delta_{\infty}$ from the MCBM being defined by the Eq.~\eqref{eq:model_MCBM_delta}.}
\label{Fig:Delta_Direct_MCBM}
\end{figure}

{\it Direct inversion for $\Delta(\omega)$.} The last thing that needs to be done, is to extract $\Delta(\omega)$ similarly to what was done in the Sec.~\ref{Subsec:Direct_Inversion} and Sec.~\ref{Subsec:Fitting_for_Delta}. As we have already seen in the main text, the direct inversion method works well in reconstructing the properties on the scale $\omega\lesssim \Delta_{\infty}$, but it completely fails on the scale $\omega\gg\Delta_{\infty}$. We can see exactly this behavior in Fig.~\ref{Fig:Delta_Direct_MCBM}.

\begin{figure}[htpb]
\includegraphics[width = 7.3 cm]{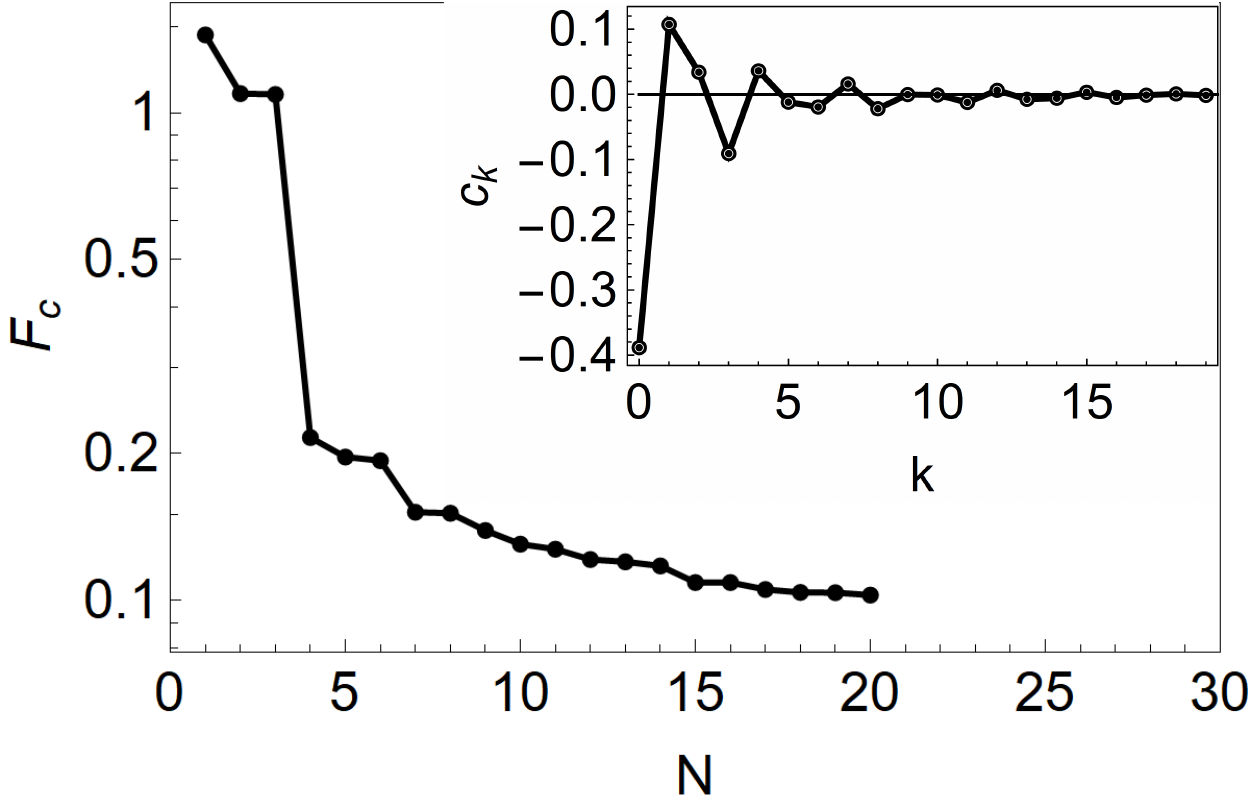}
\caption{Evolution of the cost function integral $F_c$ with the number of considered functions $\rho_k(\omega)$ and their optimal values for the coefficients $c_k$. The inset shows values of the coefficients $c_0,...,c_{19}$ minimizing $F_c$, defined in Eq.~\eqref{eq:cost_fun}. The purpose of the black line is to guide the eye.}
\label{Fig:cost_fun_compl}
\end{figure}

{\it Fitting for $\Delta(\omega)$.} The very last step that needs to be done is to extract $\Delta(\omega)$ by the method of fitting. Once again, proceeding exactly as described in Sec.~\ref{Subsec:Fitting_for_Delta}, we come to the result that shows the minimization of the cost function $F_c$ together with the rising number of considered basis functions $N$ in Fig.~\ref{Fig:cost_fun_compl}. 

For completeness, let us note, that we clearly have multiple energy scales involved in the evolution of $\widetilde{n}(\omega)$, and so also for $\Delta(\omega)$. Therefore, we assume probably the most simple and sensible combination of the parameters: $\Lambda = \Theta =  \Delta_{\infty}$ and $\omega_0 = \omega_1 = 3\Delta_{\infty}$, required by the definition of $F_c$ from Eq.~\eqref{eq:cost_fun}.

As can be noticed, beyond $N\geq 15$ the changes in the considered cost function are relatively small. This property can be understood through the behavior of the $c_k$ coefficients, which come close to $0$ for larger $N$. This behavior is presented in the inset of Fig.~\ref{Fig:cost_fun_compl}.

\begin{figure}[H]
\includegraphics[width = 7.9cm]{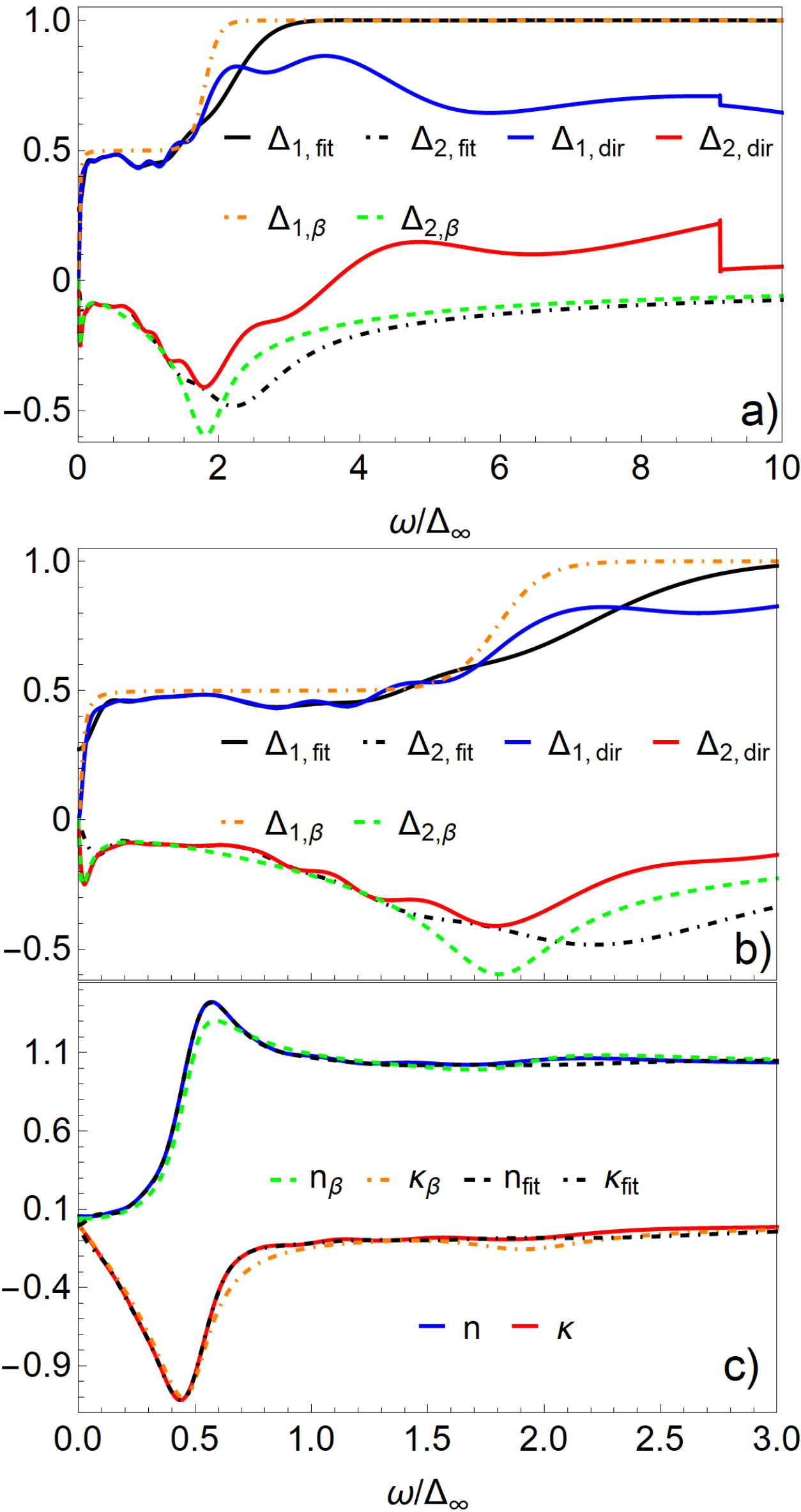}
\caption{a) Reconstructed gap functions by fitting $\Delta_{\textrm{fit}}(\omega)/\Delta_{\infty}$ (black and black-dot-dashed lines) compared with the MCBM gap function $\Delta_{\beta}(\omega)/\Delta_{\infty}$ (orange-dot-dashed and green-dashed lines). For completeness, we show also the directly inverted gap function $\Delta_{\textrm{dir}}(\omega)/\Delta_{\infty}$ (blue and red). Panel b) captures details at lower energies $0\leq\omega/\Delta_{\infty}\leq 3$. Control panel c) shows $\widetilde{n}(\omega)$ using gap functions from a) and b).}
\label{Fig:Final_MCBM}
\end{figure}

The final result of the extracted $\Delta_{\textrm{fit}}(\omega)$ is presented in Fig.~\ref{Fig:Final_MCBM}. As can be seen in Fig.~\ref{Fig:Final_MCBM}~a), the qualitative behavior of the extracted $\Delta_{\textrm{fit}}(\omega)$ corresponds to the features assumed in our challenging MCBM model. However, in Fig.~\ref{Fig:Final_MCBM}~b) we can notice that the directly extracted $\Delta_{\textrm{dir}}(\omega)$ works better assuming the energy scale $\omega \rightarrow 0$. In Fig.~\ref{Fig:Final_MCBM}~c) we can see that either $\Delta_{\textrm{fit}}(\omega)$, or $\Delta_{\textrm{dir}}(\omega)$ lead to very similar results for $n(\omega)$, which is in very good agreement with the assumed $n_{\beta}(\omega)$, of the original MCBM model.

Assuming all of the complications of the MCBM model, the reconstruction of the dos function is satisfactory. Subsequent extraction of the gap function qualitatively reveals the main physically relevant characteristics of the MCBM model gap function, which are as follows: (i) Drop of the $\Delta_1(\omega)$ at $\omega$ close to $0$ caused by the very small Dynes $\Gamma$ parameter. (ii) Step caused by the pair breaking mode at $\omega_* = 1.8\Delta_{\infty}$. Values of the gap function $\Delta_1(\omega)=0.5\Delta_{\infty}$ below $\omega_*$ and $\Delta_1(\omega) \rightarrow \Delta_{\infty}$ for $\omega\rightarrow\infty$. Further qualitative improvement can be achieved by involving more basis functions and more fitting parameters in the expansion for $\Delta_{\textrm{fit}}(\omega)$. The time-cost of such an attitude can be one of the things to address in further development of the introduced method.

\end{appendix}

\end{document}